\documentclass[a4paper,11pt]{article}
\usepackage[utf8]{inputenc}
\usepackage[T1]{fontenc}
\usepackage[english]{babel}
\usepackage[hmargin={32mm,32mm},vmargin={32mm,35mm}]{geometry}
\usepackage{amsmath}
\usepackage{amsfonts}
\usepackage{amssymb}
\usepackage[mathscr]{euscript}
\usepackage{enumerate}
\usepackage{graphicx}
\usepackage{bbm}
\usepackage{color}
\usepackage{xspace}
\usepackage{url}
\usepackage[hidelinks]{hyperref}
\usepackage{caption}
\usepackage{enumitem}
\usepackage{bookmark}
\usepackage{csquotes}
\usepackage[
	bibstyle=phys,
	biblabel=brackets,
	citestyle=numeric-comp,
	sorting=none,
	doi=false,
	eprint=true,
	pageranges=false,
	maxbibnames=10,
	backend=biber
]{biblatex}

\DeclareFieldFormat[article,inproceedings,inbook]{title}{\mkbibitalic{#1\isdot}}
\AtEveryBibitem{
	\ifentrytype{book}{
		\clearfield{series}
		\clearfield{volume}
	}{}
	\ifentrytype{inbook}{
		\clearfield{pages}
	}{}
}

\newcommand{\bra}[1]{\langle #1 |}
\newcommand{\ket}[1]{|#1\rangle}
\newcommand{\braket}[2]{\langle #1 | #2 \rangle}

\newcommand{\eg}{e.g.\@\xspace}
\newcommand{\ie}{i.e.\@\xspace}
\newcommand{\Eq}{Eq.\@\xspace}
\newcommand{\Eqs}{Eqs.\@\xspace}
\newcommand{\Fig}{Fig.\@\xspace}
\newcommand{\Figs}{Figs.\@\xspace}

\newcommand{\updown}[2]{^{#1}_{\phantom{#1}#2}}

\newcommand{\C}{\mathbbm{C}}
\newcommand{\Id}{\mathbbm{1}}

\newcommand{\Tr}{\operatorname{Tr}}
\newcommand{\D}{{\mathscr{D}}}

\newcommand{\Rt}{\,{}^{(3)}\!R}

\numberwithin{equation}{section}

\begin{document}

\begin{center}

\Large
\textbf{Time evolution of semiclassical states in the\\one-vertex model of quantum-reduced loop gravity}
 
\vspace{16pt}

\large
Ilkka Mäkinen

\normalsize

\vspace{12pt}

National Centre for Nuclear Research \\
Pasteura 7, 02-093 Warsaw, Poland
 
\vspace{8pt}
 
ilkka.makinen@ncbj.gov.pl

\end{center}

\renewcommand{\abstractname}{\vspace{-\baselineskip}}

\begin{abstract}
	\noindent We compute numerically the time evolution of simple semiclassical states describing homogeneous and isotropic spatial geometries in quantum-reduced loop gravity under a deparametrized formulation of the dynamics, in which a reference matter field is taken as a relational time variable for the dynamics of quantum states of the gravitational field. The states which we consider are defined on the Hilbert space of a spin network graph formed by a single six-valent vertex. We find that the quantum dynamics is generally in close agreement with the semiclassical effective dynamics of a homogeneous and isotropic universe throughout the range of validity of the numerical approximation. In particular, an initial state describing a contracting geometry undergoes a dynamical ``bounce'', where the contraction is halted and turned into an expansion by the quantum dynamics.
\end{abstract}

\section{Introduction}

Loop quantum gravity \cite{Ashtekar:2004eh, Rovelli:2004tv, Thiemann:2007pyv, Ashtekar:2017yom, AshtekarMa:2024} is among the major candidates for a quantum theory of gravity, providing a concrete realization of quantum gravity as a theory of quantum geometry formulated in a fully background-independent manner. The kinematical structure of loop quantum gravity is well established and is supported by a solid mathematical foundation, but gaining a satisfactory grasp of the dynamics at the level of concrete calculations remains a major challenge on the path towards a thorough understanding of the physical implications of the theory. Following the ideas originally put forward by Dirac on the quantization of constrained theories, the key object governing the dynamics of loop quantum gravity in the canonical formulation of the theory is the Hamiltonian constraint operator (see \eg \cite{Thiemann:1996aw, Lewandowski:2014hza, Assanioussi:2015gka, Yang:2015zda}), whose role is to define the physical Hilbert space of the theory and physical observables thereon.

The considerable technical difficulties encountered in the search for a sufficiently large and physically interesting set of solutions of the Hamiltonian constraint have motivated the study of alternative approaches to the problem of dynamics in loop quantum gravity. These approaches include the covariant (or spin foam) formulation of loop quantum gravity \cite{Perez:2012wv, Rovelli:2014ssa, Engle:2023qsu}, where the dynamics of the theory is encoded in transition amplitudes between quantum states of geometry, as well as the framework of deparametrized models \cite{Brown:1994py, Rovelli:1993bm, Domagala:2010bm, Husain:2011tk, Giesel:2012rb}, in which a suitable matter field is used as a physical time variable for the dynamics of the gravitational field. The dynamics is then determined by a Schrödinger-like equation, with a physical Hamiltonian operator generating time evolution of quantum states of the gravitational field with respect to the relational time defined by the reference matter field.

Our goal in this article is to examine the time evolution of simple quantum states of geometry in a deparametrized formulation of the dynamics within the setting of quantum-reduced loop gravity \cite{Alesci:2012md, Alesci:2013xd, Alesci:2013xya, Alesci:2016gub, Alesci:2018loi}. Quantum-reduced loop gravity is a model of loop quantum gravity which enjoys a significantly simplified kinematical structure, while still retaining a definite and transparent connection to the formalism of full loop quantum gravity \cite{Alesci:2013xya, Makinen:2020rda, Makinen:2023shj}. Our strategy will be to numerically compute the time evolution of a given initial state under a particular choice of the physical Hamiltonian operator. The use of numerical methods has been established in the literature of loop quantum gravity as a valid and valuable approach, which has the power to yield useful insights into the dynamics in the context of specific applications. Particularly on the covariant side of the theory, numerical computations have been employed extensively to investigate the dynamics defined by the spin foam amplitudes -- see \cite{Dona:2022yyn} and references therein. In canonical loop quantum gravity, the use of numerics to explore the dynamics has so far been restricted to a few isolated studies \cite{Assanioussi:2017tql, Kisielowski:2022wvk, Guedes:2024duc, Guedes:2024zbu}. In addition, there has recently emerged a research program which pursues the use of neural networks and deep learning methods in order to extract information about the dynamics, both in the canonical and covariant formulations \cite{Sahlmann:2024kat, Sahlmann:2024pba, Sahlmann:2025cbh}.

The specific class of quantum states whose dynamics will be investigated in this article is given by semiclassical states representing homogeneous and isotropic spatial geometries on a spin network graph consisting of a single six-valent vertex. (We assume that the graph is embedded in a spatial manifold which is topologically a three-torus, or that periodic boundary conditions are imposed, so that the graph is formed by three closed, mutually orthogonal edges, each edge both beginning and ending at the central vertex.) The Euclidean part of the physical Hamiltonian governing the dynamics of these states is obtained as a straightforward modification of a construction considered previously in \cite{Yang:2015zda, Assanioussi:2017tql, Alesci:2015wla}, while the Lorentzian part of the Hamiltonian is taken to be the scalar curvature operator introduced in \cite{Lewandowski:2021iun} and analyzed further in \cite{Lewandowski:2022xox} in the context of quantum-reduced loop gravity. The form of the resulting Hamiltonian as an operator on the Hilbert space of the one-vertex graph in quantum-reduced loop gravity has been derived in the earlier article \cite{Makinen:2024rbg}, but no concrete calculations of the dynamics generated by the operator have been performed in the literature so far.

The problem of computing the time evolution of a given initial state can be made numerically tractable by introducing a finite (but relatively large) cutoff on the spin quantum numbers labeling the basis states on the Hilbert space of the one-vertex graph. Then the action of the time evolution operator on the state can be computed by means of standard, well-established numerical algorithms for evaluating the product of the matrix exponential $e^A$ (for a sparse matrix $A$) and a vector $v$ without requiring the construction of the entire matrix $e^A$. Such an approach is naturally incapable of providing an accurate description of the dynamics over arbitrarily long time intervals, as the presence of the unphysical, arbitrarily imposed cutoff will inevitably begin disrupting the dynamics of the state at some point during its evolution. Our results nevertheless show that the approximation is powerful enough to capture a range of varied and physically relevant examples of dynamics exhibited by states on the one-vertex graph.

The material in this article is organized as follows. After the present introductory section, we review the basic kinematical framework of quantum-reduced loop gravity in section \ref{sec:qrlg}. In section \ref{sec:one-vertex} we then introduce the states and elementary operators on the Hilbert space on the one-vertex graph, and define the physical Hamiltonian operator governing the time evolution of these states under a deparametrized formulation of the dynamics. In section \ref{sec:evolution} we will outline the general approach used for the numerical computation of time evolution, and describe the semiclassical states which will be used as initial states for the computations. We also briefly discuss the semiclassical ``effective dynamics'' of a homogeneous and isotropic universe. The results of our computations are then presented in section \ref{sec:results}, where we give several examples illustrating the dynamics of semiclassical states peaked on data describing a static, an expanding or an initially contracting universe. In section \ref{sec:H_abc} we point out a certain technical observation related to the regularization of the Hamiltonian operator, which was discovered during the course of our investigations, and which we believe to be potentially interesting, even if not directly related to the main topic of this article. Finally, we summarize and discuss our results in the concluding section \ref{sec:conclusions}.

\section{Quantum-reduced loop gravity}
\label{sec:qrlg}

From the point of view of full loop quantum gravity, the Hilbert space of quantum-reduced loop gravity consists of a specific subspace of the kinematical Hilbert space of loop quantum gravity. In order to give the definition of the basis states which span the Hilbert space of quantum-reduced loop gravity, it is useful to begin by introducing some notation and conventions. Let
\begin{equation}
	\ket{jm}_i \qquad \text{($i = x, y$ or $z$)}
	\label{eq:jm_i}
\end{equation}
denote the eigenstates of the angular momentum operator $\hat J_i$; thus, the state \eqref{eq:jm_i} is an eigenstate of the operators $\hat J^2 = \hat J_x^2 + \hat J_y^2 + \hat J_z^2$ and $\hat J_i$ with eigenvalues $j(j+1)$ and $m$ respectively. For $i=z$ the states \eqref{eq:jm_i} are simply given by the standard basis $\ket{jm}$ typically considered in the quantum theory of angular momentum. The states for $i=x$ and $i=y$ can be obtained from the states $\ket{jm}$ by applying a rotation which rotates the unit vector $\hat e_z$ into $\hat e_x$ or $\hat e_y$. Note that this requirement does not uniquely determine the rotation in question. When working with the Hamiltonian constraint operator in quantum-reduced loop gravity, it is convenient to fix the ambiguity by requiring that the rotation corresponds to a cyclic permutation of the coordinate axes. (See \cite{Lewandowski:2022xox, Makinen:2024rbg} for further discussion on this point.) In what follows, we will assume that the states \eqref{eq:jm_i} are defined according to this convention.

The states $\ket{jm}_i$ having been defined, we use the notation
\begin{equation}
	D^{(j)}_{mn}(g)_i = {}_i\bra{jm}D^{(j)}(g)\ket{jn}_i
	\label{eq:D_i}
\end{equation}
to denote the matrix elements of the Wigner matrices $D^{(j)}(g)$ with respect to the basis \eqref{eq:jm_i}. Finally, we let
\begin{equation}
	\D^{(j)}_{mn}(g)_i = \sqrt{2j+1}D^{(j)}_{mn}(g)_i
	\label{eq:DD_i}
\end{equation}
denote the functions obtained from the representation matrix elements by normalizing them under the norm determined by the Haar measure on $SU(2)$ (\ie the functions \eqref{eq:DD_i} satisfy $\int_{SU(2)} dg\,\bigl|\D^{(j)}_{mn}(g)_i\bigr|^2 = 1)$. With the help of these definitions, we may now proceed to introduce the so-called reduced spin network states, which form an orthonormal basis on the Hilbert space of quantum-reduced loop gravity.

In the terminology of loop quantum gravity, a reduced spin network state is a cylindrical function, or a generalized, non-gauge invariant spin network state, which is characterized by the following properties:
\begin{itemize}
	\item The state is defined on a cubic graph embedded in the spatial manifold $\Sigma$. More precisely, we assume that there is given a fixed Cartesian background coordinate system, and the edges of the graph are aligned with the corresponding coordinate directions. The vertices of the graph are generically six-valent. The orientation of the graph is in principle arbitrary, but we fix it so that each edge is oriented towards the positive direction of the corresponding coordinate axis.
	\item Each edge of the graph carries an $SU(2)$ representation matrix, whose magnetic quantum numbers take the maximal or the minimal value (\ie $m_e = \pm j_e$) with respect to the basis \eqref{eq:jm_i}, where the label $i$ is chosen to match the direction of the given edge.
	\item It is typically assumed that all the spin quantum numbers labeling the state are large, \ie
		\begin{equation}
			j_e \gg 1
			\label{}
		\end{equation}
	for every edge $e$ of the graph.
\end{itemize}
Thus, the reduced spin network states on a fixed but arbitrary cubic graph $\Gamma_\Box$ are given by
\begin{equation}
	\prod_{e\in\Gamma_\Box} \D^{(j_e)}_{\tau_e j_e\; \sigma_e j_e}(h_e)_{i_e}
	\label{eq:reduced-state}
\end{equation}
where each label $i_e$ is assigned the value $x, y$ or $z$ according to the direction of the edge $e$, and the sign factors $\sigma_e$ and $\tau_e$ (which are associated respectively with the source and target vertices of the edge $e$) take the values $+1$ or $-1$ independently of each other. We will often refer to the space spanned by the basis states \eqref{eq:reduced-state} as the reduced Hilbert space.

In the literature of quantum-reduced loop gravity, the states \eqref{eq:reduced-state} and the resulting Hilbert space are usually introduced as the outcome of a kind of gauge-fixing procedure to a diagonal densitized triad implemented at the level of the quantum theory (see \eg \cite{Alesci:2013xd, Alesci:2013xya, Makinen:2023shj}); this procedure selects the space spanned by the states \eqref{eq:reduced-state} as a sector of the kinematical Hilbert space in which a suitable quantum analogue of the gauge conditions for diagonal triad is satisfied. Although the gauge conditions $E^a_i = 0$ for $a\neq i$ are compatible with each other at the classical level, they do not commute among themselves when represented as quantum operators, and hence satisfying them exactly in the quantum theory would be too strong of a requirement, which generally does not lead to any non-trivial solutions. Instead, the states \eqref{eq:reduced-state} are interpreted as approximate or asymptotic solutions satisfying the quantum gauge conditions approximately in the sector of large spins\footnote{
	For example, the gauge conditions constraining the triad to be diagonal can be assembled into a single master constraint operator $\hat M$. Then, applying this operator on a reduced spin network state $\ket{\Psi_0}$, one finds $\hat M\ket{\Psi_0} = {\cal O}(j^{-1/2})$ \cite{Makinen:2023shj}; hence the state becomes a more and more accurate solution of the constraint as the spins are taken larger and larger.
}. These considerations give rise to the assumption mentioned above, where the spins characterizing the reduced spin network states are usually required to be large.

We then move on to consider the question of operators on the reduced Hilbert space. The fundamental operators of quantum-reduced loop gravity enjoy a clear and definite relation with the corresponding operators of full loop quantum gravity \cite{Makinen:2020rda}. For most operators of practical interest, the action of the full theory operator on the basis states \eqref{eq:reduced-state} gives rise to a corresponding operator on the reduced Hilbert space in a natural way which will now briefly describe. Let $\ket{\Psi_0}$ denote a reduced spin network state of the form \eqref{eq:reduced-state} and consider a given operator $\hat{\cal O}$ (for example the volume operator or the Hamiltonian constraint) on the kinematical Hilbert space of full loop quantum gravity. If one computes the action of the operator $\hat{\cal O}$ on the state $\ket{\Psi_0}$ and groups the resulting terms according to their dependence on the spin quantum numbers, the result for most (although not necessarily all) operators one usually works with in practice has the following schematic structure:
\begin{equation}
	\hat{\cal O}\ket{\Psi_0} = f(j)\ket{\Phi_0} + g(j)\ket{\varphi}.
	\label{eq:O}
\end{equation}
Here $\ket{\Phi_0}$ is a normalized state belonging to the reduced Hilbert space, while $\ket{\varphi}$ is a normalized state typically not belonging to the reduced Hilbert space, and crucially, under the assumption of large spin quantum numbers, the first term on the right-hand side is large in comparison with the second one:
\begin{equation}
	f(j) \gg g(j).
	\label{}
\end{equation}
If the action of a specific operator $\hat{\cal O}$ of the full theory on the reduced spin network basis is of the form \eqref{eq:O}, one can define a corresponding operator ${}^R\hat{\cal O}$ on the reduced Hilbert space simply by discarding the second term on the right-hand side of \Eq \eqref{eq:O}:
\begin{equation}
	{}^R\hat{\cal O}\ket{\Psi_0} = f(j)\ket{\Phi_0}
	\label{eq:O^R}
\end{equation}
By definition, the action of the reduced operator ${}^R\hat{\cal O}$ preserves the reduced Hilbert space, and hence ${}^R\hat O$ is a well-defined operator for quantum-reduced loop gravity. On the other hand, from the perspective of full loop quantum gravity, the operator ${}^R\hat{\cal O}$ serves as a good approximation of the action of the full operator $\hat{\cal O}$ on the states \eqref{eq:reduced-state}. Moreover, from the practical point of view, the structure and degree of complexity of the reduced operator ${}^R\hat{\cal O}$ is generally much simpler than that of the full operator $\hat{\cal O}$. This simplified structure of the model's fundamental operators is a key advantage of quantum-reduced loop gravity when it comes to concrete applications and computations.

\section{The one-vertex model}
\label{sec:one-vertex}

\subsection{States and kinematical operators}
\label{sec:1v-kinematics}

For the remainder of this article we will focus our attention on a simple model, which is obtained from the general framework introduced in the previous section by specializing to the simplest possible cubic graph, namely one consisting of a single six-valent vertex. We assume that the graph is formed by three mutually orthogonal edges, which are embedded in a spatial manifold having the topology of a three-torus (or alternatively being characterized by periodic boundary conditions) so that each edge closes on itself, both beginning and ending at the single vertex of the graph. Furthermore, for the purposes of the work that will be presented here, it is sufficient to consider only those basis states \eqref{eq:reduced-state} in which all the sign factors $\sigma_e$ and $\tau_e$ are equal to $+1$. (This assumption will be justified in detail in section \ref{sec:numerics}.) Accordingly, we take the Hilbert space of the ``one-vertex model'' to be spanned by reduced spin network states of the form
\begin{equation}
	\ket{j_xj_yj_z} = \D^{(j_x)}_{j_xj_x}(h_{e_x})_x \D^{(j_y)}_{j_yj_y}(h_{e_y})_y \D^{(j_z)}_{j_zj_z}(h_{e_z})_z.
	\label{eq:1v-state}
\end{equation}

The elementary operators of the one-vertex model are given by reduced holonomy operators associated to the three edges of the graph, and reduced flux operators associated to the three surfaces $S^a$ $(a = x, y, z)$ such that each surface intersects the graph at the vertex and lies in the plane spanned by the other two coordinate directions. We will now briefly summarize the action of these operators on the basis states \eqref{eq:1v-state} (a detailed derivation of the results reviewed below can be found in \cite{Makinen:2020rda}). Consider first the reduced flux operators ${}^R\hat E_i(S^a)$. If the internal index $i$ takes a value different from $a$, the action of the operator on the states \eqref{eq:1v-state} gives zero. For the non-vanishing components of the reduced flux operators, we adopt the shorthand notation
\begin{equation}
	\hat p_a = {}^R\hat E_a(S^a).
	\label{}
\end{equation}
The action of these operators on the basis \eqref{eq:1v-state} is diagonal:
\begin{equation}
	\hat p_a\ket{j_xj_yj_z} = j_a\ket{j_xj_yj_z}.
	\label{}
\end{equation}
A closely related operator is the reduced volume operator. For a vertex $v$ of an arbitrary cubic graph, the reduced volume operator (corresponding to the Ashtekar--Lewandowski volume operator of the full theory \cite{Ashtekar:1997fb}) can be expressed in terms of the flux operators as
\begin{equation}
	{}^R\hat V_v = {}^R\sqrt{\bigl|\hat E_x(S^x)\hat E_y(S^y)\hat E_z(S^z)\bigr|}.
	\label{}
\end{equation}
In the context of the one-vertex model considered here, we will denote the reduced volume operator associated with the single vertex simply by $\hat v$. Like the reduced flux operators, this operator as well acts diagonally on the states \eqref{eq:1v-state}:
\begin{equation}
	\hat v\ket{j_xj_yj_z} = \sqrt{j_xj_yj_z}\ket{j_xj_yj_z}.
	\label{}
\end{equation}
Let us then consider the reduced holonomy operators. Their action on the basis states is given by
\begin{equation}
	{}^R\widehat{D^{(s)}_{mn}(h_{e_x})_x}\ket{j_xj_yj_z} = \delta_{mn}\ket{j_x + m, j_y, j_z}
	\label{eq:D^R}
\end{equation}
together with analogous equations for the edges $e_y$ and $e_z$. In other words, when the matrix elements of the holonomy are expressed in the basis $\ket{jm}_i$ matching the direction of the edge the operator is acting on, only the diagonal matrix elements of the reduced holonomy operator have a non-vanishing action on reduced spin network states. The action of the diagonal matrix elements is formally identical to a $U(1)$ multiplication law, where the $U(1)$ ``charge'' imparted on the state is determined by the magnetic quantum number (and not the spin) carried by the operator.

For the purpose of working with the Hamiltonian constraint operator, which will be introduced in the next section, it is useful to define the following symmetric operators constructed out of the reduced holonomy operator:
\begin{equation}
	\hat c_a^{(m)} = \frac{1}{2}\Bigl({}^R\widehat{D^{(s)}_{mm}(h_{e_a})_a} + {}^R\widehat{D^{(s)}_{-m, -m}(h_{e_a})_a}\Bigr)
	\label{eq:c_a}
\end{equation}
and
\begin{equation}
	\hat s_a^{(m)} = \frac{1}{2i}\Bigl({}^R\widehat{D^{(s)}_{mm}(h_{e_a})_a} - {}^R\widehat{D^{(s)}_{-m, -m}(h_{e_a})_a}\Bigr).
	\label{eq:s_a}
\end{equation}
Note that, although the expressions on the right-hand side seemingly depend on the spin $s$ in addition to the magnetic quantum number $m$, their action on the states $\ket{j_xj_yj_z}$ is in fact fully determined by the magnetic number, as shown by \Eq \eqref{eq:D^R}. Hence it is not necessary to include the spin among the labels carried by the operators $\hat c_a^{(m)}$ and $\hat s_a^{(m)}$. Indeed, their action on the states \eqref{eq:1v-state} reads
\begin{align}
	\hat c_x^{(m)}\ket{j_xj_yj_z} &= \frac{1}{2}\Bigl(\ket{j_x + m, j_y, j_z} + \ket{j_x - m, j_y, j_z}\Bigr) \\[1ex]
	\hat s_x^{(m)}\ket{j_xj_yj_z} &= \frac{1}{2i}\Bigl(\ket{j_x + m, j_y, j_z} - \ket{j_x - m, j_y, j_z}\Bigr)
	\label{}
\end{align}
and similar equations of course hold for $a=y$ and $a=z$.

\subsection{Dynamics and the physical Hamiltonian operator}
\label{sec:H_phys}

In the canonical formulation of loop quantum gravity, the central object governing the dynamics of the theory is the Hamiltonian constraint operator (see \eg \cite{Thiemann:1996aw, Lewandowski:2014hza, Assanioussi:2015gka, Yang:2015zda}). In the fully constrained approach to the dynamics, the Hamiltonian constraint is interpreted straightforwardly as a constraint operator, which determines the physical Hilbert space of the theory as the space of solutions of the constraint equation
\begin{equation}
	\hat C(N)\ket{\Psi_{\rm phys}} = 0
	\label{eq:C=0}
\end{equation}
(where $N$ is the lapse function). An alternative point of view, which can often be somewhat more manageable from the practical standpoint, is provided by the deparametrized formulation of the dynamics (see \eg \cite{Rovelli:1993bm, Domagala:2010bm, Husain:2011tk, Giesel:2012rb}). Here one considers the gravitational field coupled to a suitable reference matter field, which is usually taken to be some type of a scalar field; this matter field then serves as a physical, relational time variable, with respect to which the dynamics of the gravitational field is described. In the quantum theory, instead of the constraint equation \eqref{eq:C=0}, the dynamics of quantum states of the gravitational field is encoded in the Schrödinger equation
\begin{equation}
	i\frac{d}{dT}\ket{\Psi} = \hat H_{\rm phys}\ket{\Psi},
	\label{eq:SE}
\end{equation}
where $T$ denotes the physical time variable defined by the reference matter field, and the physical Hamiltonian $\hat H_{\rm phys}$ is an operator which is related in some way to the Hamiltonian constraint, and whose precise form depends on the type of reference matter field employed. In the case of an irrotational dust field \cite{Brown:1994py, Husain:2011tk, Swiezewski:2013jza}, the relation between the Hamiltonian constraint and the physical Hamiltonian is particularly straightforward, as one has
\begin{equation}
	\hat H_{\rm phys} = \hat C(1),
	\label{eq:H=C}
\end{equation}
\ie the physical Hamiltonian is simply the Hamiltonian constraint evaluated at lapse function $N=1$.

In what follows, we will consider a specific form of the Hamiltonian constraint operator for quantum-reduced loop gravity, which originates from a quantization of the classical functional
\begin{equation}
	C(N) = \frac{1}{\beta^2}\int d^3x\,N\biggl(\frac{\epsilon\updown{ij}{k}E^a_iE^b_jF_{ab}^k}{\sqrt{|\det E|}} + (1 + \beta^2)\sqrt{|\det E|}\Rt\biggr)
	\label{eq:C_class}
\end{equation}
on a fixed cubic graph. Here $\beta$ is the Barbero--Immirzi parameter, while $E^a_i$ and $F_{ab}^i$ are respectively the densitized triad and the curvature of the Ashtekar connection. Note that the Lorentzian part of the Hamiltonian is represented by the three-dimensional Ricci scalar $\Rt$, instead of an expression involving the extrinsic curvature $K_a^i$, which is perhaps the more well-known form of the Hamiltonian constraint in the loop quantum gravity literature. The operator in question has been previously studied in \cite{Makinen:2024rbg} both in the context of quantum-reduced loop gravity in general, as well as the one-vertex model described in section \ref{sec:one-vertex}. Accordingly, we will restrict ourselves here to giving only a concise summary of the operator, focusing on those aspects of the construction that are relevant for the work carried out in the present article, while referring the reader to \cite{Makinen:2024rbg} for a more thorough and technically comprehensive presentation.

The Euclidean part of the Hamiltonian is given by a variant of an operator considered previously in \cite{Alesci:2015wla, Yang:2015zda, Assanioussi:2017tql} in the setting of full loop quantum gravity. The loop of the holonomy regularizing the curvature of the Ashtekar connection is constructed according to a graph-preserving prescription, using a minimal rectangular loop on the cubic graph. The Lorentzian part is taken as the scalar curvature operator introduced in \cite{Lewandowski:2021iun} and examined further in \cite{Lewandowski:2022xox} in the context of quantum-reduced loop gravity. An essential technical feature of both the Euclidean and Lorentzian operators is the method used to deal with the quantization of the inverse volume element $1/\sqrt{|\det E|}$ in the Euclidean part of the constraint \eqref{eq:C_class}, as well as analogous factors that are encountered in the quantization of the Ricci scalar. Instead of applying the technique known as Thiemann's trick \cite{Thiemann:1996aw} (where the classical expression \eqref{eq:C_class} is rewritten using identities that replace the problematic factors with Poisson brackets that result in a non-singular operator upon quantization), the factors $1/\sqrt{|\det E|}$ are quantized directly by using the so-called Tikhonov regularization of the inverse volume operator. Given the local volume operator $\hat V_v$ at a vertex $v$, the regularized inverse volume operator is defined by the expression
\begin{equation}
	\widehat{{\cal V}_v^{-1}} = \lim_{\epsilon\to 0} \frac{\hat V_v}{\hat V_v^2 + \epsilon^2}.
	\label{eq:V^-1}
\end{equation}
An alternative, but equivalent definition of the operator $\widehat{{\cal V}_v^{-1}}$ can be given by specifying its spectral decomposition as $\widehat{{\cal V}_v^{-1}} = \sum_{\lambda\neq 0}\lambda^{-1}\ket{\lambda}\bra{\lambda}$, where $\{\ket{\lambda}\}$ is a complete set of eigenstates of the operator $\hat V_v$ with corresponding eigenvalues $\lambda$.

The action of the Hamiltonian constraint operator on the one-vertex states \eqref{eq:1v-state} has been derived in \cite{Makinen:2024rbg} under a particular, non-symmetric choice of factor ordering. Separating the full constraint operator into its Euclidean and Lorentzian parts by writing 
\begin{equation}
	{}^R\hat C(N) = \frac{1}{\beta^2}{}^R\hat C_E(N) + \frac{1 + \beta^2}{\beta^2}{}^R\hat C_L(N),
	\label{eq:C^R}
\end{equation}
the result reads
\begin{equation}
	{}^R\hat C_E(N)\ket{j_xj_yj_z} = -N(v)\Biggl[\sqrt{\frac{j_xj_y}{j_z}}\hat s_x^{(1)}\hat s_y^{(1)} \; + \; \text{cycl.~perm.}\Biggr]\ket{j_xj_yj_z}
	\label{eq:C_E^R}
\end{equation}
for the Euclidean part, and
\begin{equation}
	{}^R\hat C_L(N)\ket{j_xj_yj_z} = -16N(v)\Biggl[\frac{j_x^{3/2}}{\sqrt{j_yj_z}}\bigl(\hat s_x^{(1/2)}\bigr)^4 \; + \; \text{cycl.~perm.}\Biggr]\ket{j_xj_yj_z}
	\label{eq:C_L^R}
\end{equation}
for the Lorentzian part. Here $\hat s_a^{(m)}$ is the operator defined by \Eq \eqref{eq:s_a}, and ``cycl.~perm.'' in each equation stands for two additional terms, which are obtained from the term explicitly shown via cyclic permutations of the labels $x$, $y$ and $z$. The factors of $1/\sqrt{j_a}$ in \Eqs \eqref{eq:C_E^R} and \eqref{eq:C_L^R} arise from the Tikhonov regularization of the inverse volume, and hence constitute a mild abuse of notation: If any spin $j_a = 0$ in the state $\ket{j_xj_yj_z}$, the corresponding factor $1/\sqrt{j_a}$ should be understood as being equal to zero.

We now wish to modify the operator defined by \Eqs \eqref{eq:C^R}--\eqref{eq:C_L^R} in order to construct a physical Hamiltonian in the way indicated by \Eq \eqref{eq:H=C} for a deparametrized formulation of the dynamics, in which an irrotational dust field is used as a physical time variable for the quantum dynamics of the gravitational field. That is, the physical Hamiltonian will have the form
\begin{equation}
	\hat H_{\rm phys} = \frac{1}{\beta^2}\hat H_E + \frac{1 + \beta^2}{\beta^2}\hat H_L
	\label{eq:H_phys}
\end{equation}
where the operators $\hat H_E$ and $\hat H_L$ are obtained from \Eqs \eqref{eq:C_E^R} and \eqref{eq:C_L^R} by setting $N = 1$ and adjusting the factor ordering in a way that results in a symmetric operator. For the Euclidean part, we will adopt a factor ordering that amounts to distributing the factor $\sqrt{j_xj_y/j_z}$ symmetrically on both sides of the holonomy operators in \Eq \eqref{eq:C_E^R}. To spell this out explicitly, it is helpful to introduce the operator
\begin{equation}
	\hat X_a = \sqrt{\frac{\hat p_b\hat p_c}{\hat p_a}} = \frac{\hat v}{\hat p_a}
	\label{eq:X_a}
\end{equation}
where in the intermediate step $(a, b, c)$ denotes any permutation of $(x, y, z)$. Then a symmetric factor ordering of the Euclidean operator is given by
\begin{equation}
	\hat H_E = -\hat X_z^{1/2}\hat s_x^{(1)}\hat s_y^{(1)}\hat X_z^{1/2} - \hat X_y^{1/2}\hat s_x^{(1)}\hat s_z^{(1)}\hat X_y^{1/2} - \hat X_x^{1/2}\hat s_y^{(1)}\hat s_z^{(1)}\hat X_x^{1/2}.
	\label{eq:H_E}
\end{equation}
Note that the operators $\hat s^{(1)}_a$ commute among each other, so there is no question about their relative ordering in the above equation. The operator $1/\hat p_a$ in \Eq \eqref{eq:X_a} should again be understood in the sense of the Tikhonov prescription, \ie
\begin{equation}
	\frac{1}{\hat p_a}\ket{j_xj_yj_z} = 0 \qquad \text{if} \quad j_a = 0.
	\label{}
\end{equation}
In order to specify a factor ordering for the Lorentzian part, we define another auxiliary operator as
\begin{equation}
	\hat Y_a = \frac{\hat p_a^{3/2}}{\sqrt{\hat p_b\hat p_c}} = \frac{\hat p_a^2}{\hat v}.
	\label{}
\end{equation}
(The remark made above regarding the operator $1/\hat p_a$ naturally applies also to the operator $1/\hat v$ here.) We then divide each factor $j_a^{3/2}/\sqrt{j_bj_c}$ in \Eq \eqref{eq:C_L^R} into four factors of $\hat Y_a^{1/4}$ and distribute them around the holonomy operators as follows:
\begin{equation}
	\hat H_L = -16\sum_{a = x,y,z}\hat Y_a^{1/4}\bigl(\hat s_a^{(1/2)}\bigr)^2\hat Y_a^{1/2}\bigl(\hat s_a^{(1/2)}\bigr)^2\hat Y_a^{1/4}.
	\label{eq:H_L}
\end{equation}
Note that each term above is the square of the operator $\hat Y_a^{1/4}\bigl(\hat s_a^{(1/2)}\bigr)^2\hat Y_a^{1/4}$, which is itself symmetric. Similar factor orderings have been utilized previously for the Hamiltonian in models of loop quantum cosmology and LQC-like models of spherically symmetric spacetimes, see \eg \cite{Martin-Benito:2009htq, MenaMarugan:2011me, Cafaro:2026twe}. Note also that the factor $\bigl(\hat s^{(1/2)}_a\bigr)^2$ can be written as
\begin{equation}
	\bigl(\hat s^{(1/2)}_a\bigr)^2 = \frac{1}{2}\bigl(\Id - \hat c^{(1)}_a\bigr),
	\label{}
\end{equation}
which makes it explicit that the operator \eqref{eq:H_L} shifts the spins in integer steps when acting on the state $\ket{j_xj_yj_z}$. A further property, which is a consequence of the factor ordering chosen in \Eqs \eqref{eq:H_E} and \eqref{eq:H_L}, is that the matrix element of the physical Hamiltonian between the basis states $\ket{j_xj_yj_z}$ vanishes if any spin is equal to zero in the initial or the final state:
\begin{equation}
	\bra{k_xk_yk_z}\hat H_{\rm phys}\ket{j_xj_yj_z} = 0 \qquad \text{if any $j_a = 0$ or $k_a = 0$}.
	\label{eq:H_kj=0}
\end{equation}
This property turns out to be useful for the computations to which we will now turn our attention.

\section{Time evolution of semiclassical states}
\label{sec:evolution}

\subsection{The setup for numerical computations}
\label{sec:numerics}

In this section we will outline a general strategy towards numerically computing the time evolution of quantum states on the Hilbert space of the one-vertex model in a deparametrized formulation, where the physical Hamiltonian is given by the operator introduced in the previous section. The dynamics of a given state vector is then governed by the Schrödinger equation
\begin{equation}
	i\frac{d}{dT}\ket{\psi(T)} = \hat H_{\rm phys}\ket{\psi(T)}.
	\label{}
\end{equation}
Since the physical Hamiltonian is a time-independent operator, the evolution of an initial state $\ket{\psi_0}$ is given by
\begin{equation}
	\ket{\psi(T)} = e^{-i\hat H_{\rm phys}T}\ket{\psi_0}.
	\label{eq:psiT}
\end{equation}
If the eigenvalues and eigenstates of the operator $\hat H_{\rm phys}$ were known, the evolution operator $e^{-i\hat H_{\rm phys}T}$ could be constructed through its spectral decomposition. However, at least for the time being, we do not have access to the spectrum of the physical Hamiltonian (nor to the spectra of the Euclidean or Lorentzian operators $\hat H_E$ and $\hat H_L$ individually). In this circumstance, the most straightforward way to make progress is by introducing a suitable truncation of the Hilbert space of the model, such that the task of evaluating the expression \eqref{eq:psiT} is converted into a problem on a finite-dimensional Hilbert space.

In order to motivate the specific truncation that will be employed in our computations, let us first observe that the action of the operator $\hat H_{\rm phys}$ on the states $\ket{j_xj_yj_z}$ shifts the quantum numbers of the state in integer steps. This implies that the subspace spanned by the basis states in which all the quantum numbers take integer values is preserved by the action of the physical Hamiltonian.\footnote{\label{fn:subspaces}
	More generally, the entire Hilbert space of the one-vertex model splits into eight orthogonal subspaces, according to whether each quantum number $j_x$, $j_y$ and $j_z$ is an integer or a half-integer; these subspaces are dynamically decoupled, each of then being preserved by the operator $\hat H_{\rm phys}$.
}
Furthermore, within the space characterized by integer quantum numbers, the matrix element of the physical Hamiltonian between two basis states vanishes whenever all the quantum numbers in the initial state are positive while any quantum number in the final state is negative or equal to zero. For the Euclidean part of the Hamiltonian, whose action on the basis states shifts any quantum number at most by one unit, this follows directly from \Eq \eqref{eq:H_kj=0}. In contrast, the action of the Lorentzian part may generally shift a quantum number by two units; however, the factor ordering chosen in \Eq \eqref{eq:H_L} guarantees that the matrix element $\bra{k_xk_yk_z}\hat H_L\ket{j_xj_yj_z}$ vanishes if any quantum number in the initial state is equal to $1$ and the corresponding quantum number in the final state is equal to $-1$.

The argument given above identifies the space spanned by the states $\ket{j_xj_yj_z}$, where all the quantum numbers are positive integers, as a subspace preserved under the action of the Hamiltonian. Hence it also justifies the assumption introduced in the opening paragraph of section \ref{sec:1v-kinematics}, where the sign factors $\sigma_e$ and $\tau_e$ in the general expression \eqref{eq:reduced-state} for the basis states were fixed to take the value $+1$. Finally, we truncate the space we have obtained by arbitrarily imposing an upper cutoff $j_{\rm max}$ on the quantum numbers labeling the basis states. Thus, the truncated Hilbert space is given by
\begin{equation}
	{\cal H}_{j_{\rm max}} = {\rm span}\Bigl(\ket{j_xj_yj_z} : j_a \in \{1, 2, \dots, j_{\rm max}\} \Bigr).
	\label{eq:H_jmax}
\end{equation}
Generally one would expect higher and higher values of the cutoff to presumably provide better and better approximations of the true quantum dynamics of the one-vertex model. In practice, the appropriate choice of the cutoff is largely dictated by the amount of computational power available. In the computations that will be presented here, the value $j_{\rm max} = 200$ was used, giving rise to a truncated Hilbert space of dimension $j_{\rm max}^3 = 8\times 10^6$.

Within the truncated Hilbert space, the time evolution of a state vector is computed by repeatedly evaluating
\begin{equation}
	\ket{\psi(T + \Delta T)} = e^{-i\hat H_{\rm phys}\Delta T}\ket{\psi(T)}
	\label{eq:psi_t+dt}
\end{equation}
for a fixed, sufficiently small time step $\Delta T$. The action of the evolution operator on the state vector is computed by means of numerical libraries, which compute the product $e^A v$ for a vector $v$ and a sparse matrix $A$ without having to form the entire matrix representing the exponential $e^A$. In order to assess the reliability of the numerics, we prepared two independent computer programs, one written in Python and the other written in Julia, which utilize different numerical algorithms for the computation of the action of the matrix exponential. The Python program uses the function \texttt{expm\_multiply} from the scientific computing library SciPy, which applies an algorithm consisting of a scaling and squaring method combined with a truncated Taylor expansion \cite{AlMohy:2010}, while the Julia program uses the function \texttt{expv} from the package \texttt{ExponentialUtilities.jl}, which is based on Krylov subspace methods. For a large portion of our calculations, the time evolution of a given initial state was computed using both programs, and in all cases the results were in very close agreement with each other. The source code of the programs used to perform the calculations is made available at \url{https://github.com/imakinen/QRLG-evolution}.

At each step of the computation, after the state vector has been evolved according to \Eq \eqref{eq:psi_t+dt}, we evaluate the expectation value
\begin{equation}
	\langle \hat{\cal O}(T) \rangle = \bra{\psi(T)}\hat{\cal O}\ket{\psi(T)}
	\label{}
\end{equation}
and the dispersion, or standard deviation
\begin{equation}
	\Delta{\cal O}(T) = \sqrt{\langle \hat{\cal O}^2(T)\rangle - \langle \hat{\cal O}(T)\rangle^2}
	\label{}
\end{equation}
for certain observables, including the volume operator $\hat v$ and the symmetric holonomy operators $\hat c^{(1)}_{a}$ and $\hat s^{(1)}_a$. Additionally, as a means of estimating the range of validity of the computation, we keep track of the ``occupation number'' of the basis states in which any quantum number is equal to the cutoff value $j_{\rm max}$. Expanding the state vector at time $T$ in the basis $\ket{j_xj_yj_z}$,
\begin{equation}
	\ket{\psi(T)} = \sum_{j_xj_yj_z} \alpha_{j_xj_yj_z}(T)\ket{j_xj_yj_z},
	\label{}
\end{equation}
we can define the quantity
\begin{equation}
	n_{\rm max}(T) \; = \sum_{\text{any} \; j_a=j_{\rm max}} |\alpha_{j_xj_yj_z}(T)|^2
	\label{}
\end{equation}
whose value can be taken as a rough indicator of the extent to which the presence of the artificially imposed cutoff is affecting the accuracy of the computation. As long as $n_{\rm max}$ remains small, it is reasonable to expect that the calculation performed within the truncated Hilbert space provides a good approximation of the true time evolution of the initial state, whereas a large value of $n_{\rm max}$ indicates that the computation has exceeded its range of validity and can no longer be considered as an accurate description of the dynamics. We also keep track of the quantity
\begin{equation}
	n_{\rm min}(T) \; = \sum_{\text{any} \; j_a=1} |\alpha_{j_xj_yj_z}(T)|^2
	\label{}
\end{equation}
even though the lower limit $j_{\rm min} = 1$ is achieved through a suitable factor ordering of the Hamiltonian and hence does not represent an artificial, unphysical cutoff.

In our computations we will consider separately the Euclidean model, where the physical Hamiltonian is taken to be $\hat H_{\rm phys} = \hat H_E$, and the Lorentzian model, in which the Hamiltonian is given by the full expression \eqref{eq:H_phys}. In the case of the Euclidean model only, the amount of computational resources required can be somewhat reduced by using the observation that the action of the operator $\hat H_E$ on the state $\ket{j_xj_yj_z}$ preserves the parity (even/odd) of the sum $j_x + j_y + j_z$. Let ${\cal H}_+$ and ${\cal H}_-$ denote the subspaces of the truncated Hilbert space \eqref{eq:H_jmax} in which the sum of the quantum numbers is respectively even or odd, and decompose the state vector $\ket{\psi(T)}$ into its projections onto these subspaces as
\begin{equation}
	\ket{\psi(T)} = \ket{\psi_+(T)} + \ket{\psi_-(T)}.
	\label{}
\end{equation}
Then the state can be evolved by computing separately the products
\begin{equation}
	\ket{\psi_+(T + \Delta T)} = e^{-i\hat H_E^{(+)}\Delta T}\ket{\psi_+(T)}
	\label{}
\end{equation}
and
\begin{equation}
	\ket{\psi_-(T + \Delta T)} = e^{-i\hat H_E^{(-)}\Delta T}\ket{\psi_-(T)},
	\label{}
\end{equation}
where $\hat H_E^{(+)}$ and $\hat H_E^{(-)}$ denote the projections of the Hamiltonian onto the subspaces of even and odd parity. When performed in this way, the computation requires a smaller amount of memory compared to evaluating the product \eqref{eq:psi_t+dt} on the entire Hilbert space ${\cal H}_{j_{\rm max}}$, while there is no significant difference in the total computation time. We emphasize again that this observation applies only in the case of the Euclidean model; the subspaces ${\cal H}_\pm$ are not preserved by the action of the Lorentzian operator $\hat H_L$.

\subsection{The choice of initial states}

Our goal is to apply the general machinery described in the previous section to study the dynamics of a certain class of quantum states which enjoy a transparent physical interpretation, namely semiclassical states on the one-vertex graph. The so-called complexifier coherent states, or heat-kernel coherent states, were introduced into loop quantum gravity by Thiemann \cite{Thiemann:2000bw, Thiemann:2002vj} on the basis of earlier related work by Hall \cite{Hall:1994}; these states have been further studied \eg in \cite{Thiemann:2000ca, Thiemann:2000bx, Thiemann:2000by, Bahr:2007xn, Bianchi:2009ky}. A coherent state on the Hilbert space associated with a single spin network edge is defined by the expression
\begin{equation}
	\psi_{h_0}(h_e) = \sum_j (2j+1) e^{-tj(j+1)/2} \chi^{(j)}(h_0^{-1}h_e).
	\label{eq:psi_h0}
\end{equation}
Here the group element $h_0 \in SL(2, \C)$ parametrizes the point of the classical holonomy-flux phase space on which the coherent state is peaked, $\chi^{(j)}(h) = \Tr D^{(j)}(h)$ denotes the trace (character) in the spin-$j$ irreducible representation, and the ``heat kernel time'' $t$ is a parameter which controls the semiclassical properties of the state.

For the purposes of the present work, we will be interested in states which are peaked on a diagonal holonomy and flux. 
Considering a given edge $e$ of the one-vertex graph, such states are given by\footnote{
	This is obtained from \Eq \eqref{eq:psi_h0} by decomposing the $SL(2, \C)$ group element into a ``rotation'' and a ``boost'' as $h_0 = g_0 e^{\vec p_0\cdot\vec\sigma/2}$, where $g_0 \in SU(2)$, and taking the group element $g_0 = e^{ic_0\sigma_{i_e}}$ to be generated by the single Pauli matrix $\sigma_{i_e}$, while the vector $\vec p_0$ is chosen to be aligned with the direction of the edge.
}
\begin{equation}
	\ket{j_0, c_0}_{i_e} = \sum_j \sqrt{2j + 1} e^{-t(j - j_0)^2/2}e^{-ic_0j}\D^{(j)}_{jj}(h_e)_{i_e}.
	\label{eq:ket_j0c0}
\end{equation}
On the entire one-vertex graph, we may then construct the states
\begin{equation}
	\ket{\psi_{j_0, c_0}} = {\cal N} \ket{j_0, c_0}_x \ket{j_0, c_0}_y \ket{j_0, c_0}_z
	\label{eq:psi_j0c0}
\end{equation}
where the normalization factor ${\cal N} = \bigl|\braket{j_0, c_0}{j_0, c_0}\bigr|^{-3/2}$ has to be inserted because the states \eqref{eq:ket_j0c0} are in general not normalized. The states \eqref{eq:psi_j0c0} -- or more precisely, their projection onto the truncated Hilbert space ${\cal H}_{j_{\rm max}}$ introduced in the previous section -- are taken as the initial states whose time evolution we will compute. Since the same values of $j_0$ and $c_0$ are associated to each edge, these states can be interpreted as describing a homogeneous and isotropic spatial geometry on the one-vertex graph. More generally one could consider anisotropic states, in which each state in \Eq \eqref{eq:psi_j0c0} would be labeled by different values of $j_0$ and $c_0$; however, in order to keep the scope of the computations to be performed at a manageable level, we have so far studied only the dynamics of the isotropic states \eqref{eq:psi_j0c0}.

The elementary operators of the one-vertex model have the following expectation values with respect to the state \eqref{eq:ket_j0c0}:
\begin{align}
	{}_{i_e}\bra{j_0, c_0} \hat p_{i_e} \ket{j_0, c_0}_{i_e} &\simeq j_0 + \frac{1}{2} \label{eq:p_0} \\[1ex]
	{}_{i_e}\bra{j_0, c_0} \hat s_{i_e}^{(m)} \ket{j_0, c_0}_{i_e} &\simeq \sin mc_0 \\[1ex]
	{}_{i_e}\bra{j_0, c_0} \hat c_{i_e}^{(m)} \ket{j_0, c_0}_{i_e} &\simeq \cos mc_0
\end{align}
\ie the parameters $c_0$ and $j_0$ are directly related to the expectation values of the holonomy and flux operators in the state. Good semiclassical states at the kinematical level are of course not only characterized by having certain expectation values of the elementary observables, but also by being well peaked around the expectation values (as measured by the dispersion $\delta{\cal O} = \sqrt{\langle {\cal O}^2\rangle - \langle {\cal O}\rangle^2}$). The peakedness properties of the coherent states \eqref{eq:psi_h0} have been analyzed extensively in \cite{Thiemann:2000ca}. The state is sharply peaked on the holonomy provided that the parameter $t$ is small, while peakedness with respect to the flux requires that the width of the Gaussian profile in \Eq \eqref{eq:ket_j0c0} is small relative to the location of the peak, \ie $1/\sqrt{t} \ll j_0$.

In \cite{Bianchi:2009ky}, the value $t = 1/\sqrt{j_0}$ was identified as a choice which satisfies both of the above requirements in the regime of large $j_0$, and hence guarantees good semiclassical properties of the state. While this is certainly true as far as the kinematical peakedness properties of the state are concerned, our calculations suggest that this choice is not an optimal one when it comes to the semiclassical properties of the state at the dynamical level. The scaling $t \sim 1/\sqrt{j_0}$ seems to produce states which are initially very sharply peaked on the flux operators, but spread rather quickly as they evolve, and lose their good peakedness properties at a relatively early stage of the evolution. In contrast, the choice $t \sim 1/j_0$ gives rise to states in which the spreading of the ``wave packet'' is not as rapid, and the state remains well peaked over a longer time interval through its evolution. This behavior is illustrated by the first example presented in section \ref{sec:H_E}. For the remainder of the computations, the value of the parameter $t$ will be fixed to $t = 1/2j_0$.

\subsection{Effective dynamics}
\label{sec:effective}

The quantum dynamics of the states \eqref{eq:psi_j0c0} will be compared against the effective semiclassical dynamics of a homogeneous and isotropic universe. In the context of loop quantum cosmology (and related symmetry-reduced models), the concept of effective dynamics (see \eg \cite{Ashtekar:2006wn, Taveras:2008ke, Ashtekar:2011ni, Dapor:2017rwv}) refers to the dynamics generated on the classical phase space by an effective semiclassical Hamiltonian function, whose form is motivated by considerations coming from the quantum theory, and which is intended to capture at least some relevant aspects of the quantum dynamics of the model.

For a homogeneous and isotropic spatial geometry, the Ashtekar connection and the densitized triad take the form
\begin{equation}
	A_a^i = c(t)\delta_a^i, \qquad E^a_i = p(t)\delta^a_i.
	\label{}
\end{equation}
The classical phase space is parametrized by the variables $c$ and $p$, which are canonically conjugate, satisfying the Poisson bracket
\begin{equation}
	\{c, p\} = \frac{\beta}{3}
	\label{}
\end{equation}
(in units where $8\pi G = 1$). If we consider a matter field coupled to a gravitational field, the effective Hamiltonian constraint for the system is given by an expression of the type
\begin{equation}
	H_{\rm eff} = H_{\rm gr}(c, p) + H_{\rm matter}.
	\label{}
\end{equation}
Then the effective equation of motion for a function $F(c, p)$ depending on the gravitational variables reads
\begin{equation}
	\dot F(c, p) = \{F, H_{\rm eff}\},
	\label{}
\end{equation}
where $\dot F = dF/dt$ denotes the derivative with respect to the coordinate time $t$. If the matter field is an irrotational dust field, then the equation of motion with respect to the physical time $T$ defined by the dust field is
\begin{equation}
	\frac{d}{dT}F(c, p) = \{F, H_{\rm gr}(c, p)\}.
	\label{eq:dFdT}
\end{equation}
The effective Hamiltonian appearing in \Eq \eqref{eq:dFdT} will generally have the form
\begin{equation}
	H_{\rm gr} = H_{\rm eff}^E + H_{\rm eff}^L.
	\label{}
\end{equation}
The expression representing the Euclidean part of the effective Hamiltonian is well known in the literature, and is given by
\begin{equation}
	H_{\rm eff}^E = -\frac{3}{\beta^2}\sqrt p\frac{\sin^2\mu c}{\mu^2}.
	\label{eq:H_eff^E}
\end{equation}
Here $\mu$ is a parameter, whose purpose is to encode the fact that spatial geometry is discrete according to loop quantum gravity; the classical Hamiltonian constraint $H = -(3/\beta^2)\sqrt{p}c^2$ of a homogeneous and isotropic universe is recovered from \eqref{eq:H_eff^E} in the limit $\mu\to 0$. While the expression \eqref{eq:H_eff^E} initially originates from loop quantum cosmology, from the perspective of loop quantum gravity it has been shown that the same expression can be recovered as the expectation value of the (Euclidean) Hamiltonian constraint operator in a suitable semiclassical state describing a homogeneous and isotropic geometry \cite{Dapor:2017gdk, Zhang:2021qul}. No analogous calculation is available at the moment for the scalar curvature operator which gives rise to the Lorentzian part of the physical Hamiltonian \eqref{eq:H_phys}. However, by appealing to the formal similarity between the Euclidean operator \eqref{eq:C_E^R} and the effective Hamiltonian \eqref{eq:H_eff^E}, and extending the analogy to the Lorentzian operator \eqref{eq:C_L^R}, one can propose the expression
\begin{equation}
	H_{\rm eff}^L = -48\frac{1 + \beta^2}{\beta^2}\sqrt p\frac{\sin^4(\mu c/2)}{\mu^2}
	\label{eq:H_eff^L}
\end{equation}
as a tentative, preliminary conjecture (see \cite{Makinen:2024rbg} for further discussion). When considering the quantum dynamics of the Lorentzian model with the full physical Hamiltonian \eqref{eq:H_phys}, we will use the expression \eqref{eq:H_eff^L} as the Lorentzian part of the effective Hamiltonian which generates the corresponding semiclassical effective dynamics.

\section{Results}
\label{sec:results}

In this section we present a selection of numerical results on the time evolution of different kinds of semiclassical initial states. First, in section \ref{sec:H_E}, we consider the Euclidean model, for which the physical Hamiltonian is given by
\begin{equation}
	\hat H_{\rm phys} = \hat H_E.
	\label{}
\end{equation}
The Lorentzian model, with the full physical Hamiltonian
\begin{equation}
	\hat H_{\rm phys} = \frac{1}{\beta^2}\hat H_E + \frac{1 + \beta^2}{\beta^2}\hat H_L,
	\label{}
\end{equation}
is then considered in section \ref{sec:H_L}. In the plots that follow, the expectation value and the dispersion of the volume operator $\hat v$ and the symmetric holonomy operators $\hat c_x^{(1)}$ and $\hat s_x^{(1)}$ in the evolving quantum state are displayed as a function of the physical time $T$ defined by the irrotational dust field\footnote{
	In natural units the dust field $T$ has dimensions of length, or equivalently time. In this article we work in units where $8\pi G = 1$; hence the unit of time in the plots that will be presented below is $\sqrt{8\pi}t_{\rm P}$, where $t_{\rm P} = \sqrt{\hbar G}$ is the Planck time.
}. The quantities $n_{\rm max}$ and $n_{\rm min}$, which represent the occupation number of the basis states in which any spin is equal to the cutoff value $j_{\rm max}$ or the minimum value $j_{\rm min} = 1$, are also plotted as a function of time. The occupation number $n_{\rm max}$ serves as a rough measure of the range of validity of the computation: A non-negligible value of $n_{\rm max}$ indicates that the presence of the artificially imposed cutoff has began to affect the results of the computation, after which point one can no longer expect the numerics to accurately describe the true dynamics of the state. 

The initial states for the computations are taken to be the coherent states $\ket{\psi_{j_0c_0}}$ defined by \Eq \eqref{eq:psi_j0c0}, with varying values of the parameters $c_0$ and $j_0$. For a state characterized by a given $j_0$, the value of the semiclassicality parameter $t$ is set to
\begin{equation}
	t = \frac{1}{2j_0}
	\label{}
\end{equation}
with the exception of the first example given in section \ref{sec:H_E}, where this value is compared against the alternative choice $t = 1/2\sqrt{j_0}$. All the computations are performed using the cutoff value
\begin{equation}
	j_{\rm max} = 200.
	\label{}
\end{equation}
The quantum dynamics described by the expectation values of the geometric operators is contrasted with the semiclassical effective dynamics generated by the effective Hamiltonian discussed in section \ref{sec:effective}. The semiclassical trajectory defined by the functions
\begin{equation}
	c_{\rm eff}(T) \qquad \text{and} \qquad p_{\rm eff}(T)
	\label{}
\end{equation}
is determined by numerically integrating the equations of motion \eqref{eq:dFdT}. In accordance with \Eq \eqref{eq:p_0} for the expectation value of the flux operator in the initial state $\ket{\psi_{j_0c_0}}$, the initial value $p_{\rm eff}(0)$ is set to $j_0 + 1/2$. The expectation values of the volume operator and the symmetric holonomy operators will then be compared with the quantities $v_{\rm eff}(T) = p_{\rm eff}(T)^{3/2}$, and $\cos c_{\rm eff}(T)$ and $\sin c_{\rm eff}(T)$ respectively.

Before proceeding to discuss the results, let us briefly comment on the choice to restrict our attention to the subspace in which all the quantum numbers $j_x$, $j_y$ and $j_z$ are integers, and to disregard the rest of the dynamically stable subspaces mentioned in footnote \ref{fn:subspaces}, where some (or all) of the quantum numbers are half-integers. As long as the quantum state remains supported on relatively large quantum numbers throughout its evolution, and the basis states with quantum numbers near zero remain essentially unoccupied, it seems unlikely that there would be any significant differences between the subspaces in terms of their dynamical behavior, considering that the matrix elements of the Hamiltonian are given by simple, well-behaved algebraic expressions in terms of the quantum numbers. We would expect the situation to be different in cases where the quantum evolution enters the regime of low quantum numbers, due to the fact that a half-integer quantum number can cross from positive to negative values under the Hamiltonian considered here, while an integer quantum number cannot. However, none of the examples presented below are of the latter kind: in all of them the occupation number $n_{\rm min}$ remains extremely small through the entire time interval over which the numerical computation is expected to be accurate.

\subsection{Euclidean model}
\label{sec:H_E}

\subsubsection*{Static universe}

\begin{figure}[p]
	\centering
	\includegraphics[width=\textwidth]{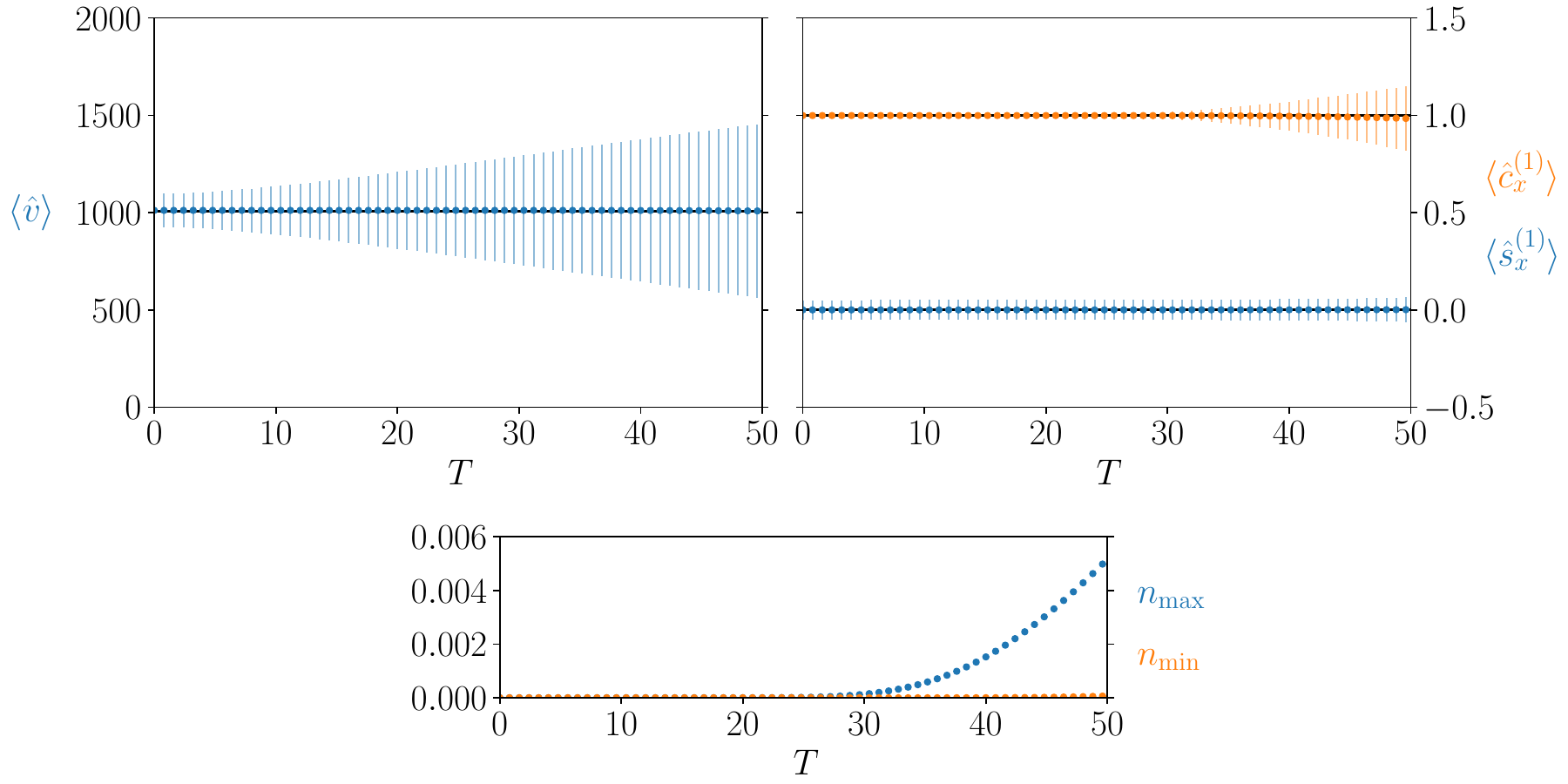}
	\caption{Dynamics of a state corresponding to a classically static configuration in the Euclidean model. The parameters describing the initial state are given by $j_0 = 100$, $c_0 = 0$. The semiclassicality parameter is set to the value $t = 1/2j_0 = 1/200$. The expectation values of the quantum observables closely follow the effective semiclassical trajectory, and the occupation number $n_{\rm max}$ remains small throughout most of the time interval.}
	\label{fig:H_E-static-1}
\end{figure}

\begin{figure}[p]
	\centering
	\includegraphics[width=\textwidth]{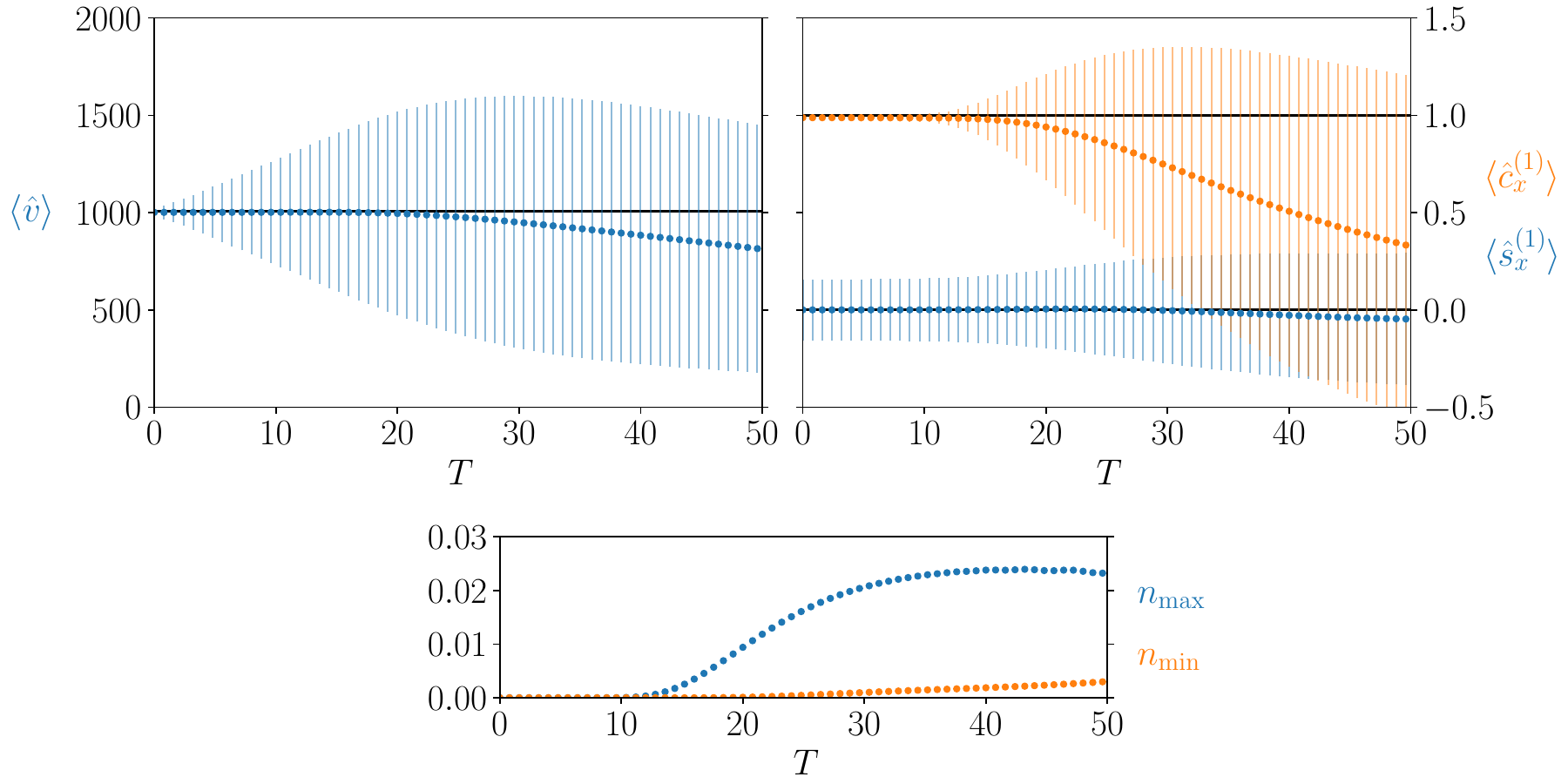}
	\caption{Another example of a state in the Euclidean model corresponding to a classically static configuration. The initial state is again described by $j_0 = 100$, $c_0 = 0$ but the semiclassicality parameter has the value $t = 1/2\sqrt{j_0} = 1/20$. The state is initially sharply peaked on the volume operator but spreads rapidly. The occupation number $n_{\rm max}$ reaches a substantial value at a relatively early stage of the evolution, indicating the end of validity of the numerical approximation.}
	\label{fig:H_E-static-2}
\end{figure}

As our first example, we consider initial data describing at the classical level a static, dynamically unchanging universe. This behavior corresponds to the initial value $c_0 = 0$ for the connection. We take the initial state to be defined by the values $j_0 = 100$ and $c_0 = 0$, and examine the two choices $t = 1/2j_0 = 1/200$ and $t = 1/2\sqrt{j_0} = 1/20$ for the semiclassicality parameter $t$. Plots illustrating the dynamics are displayed in \Figs \ref{fig:H_E-static-1} and \ref{fig:H_E-static-2} respectively for the two states. In each plot, the solid dots denote the expectation value, while the vertical bars centered around the dots indicate the dispersion $\Delta{\cal O} = \sqrt{\langle {\cal O}^2\rangle - \langle {\cal O}\rangle^2}$ of the observable in question. The effective semiclassical trajectory corresponding to each observable is plotted as a solid line.

In \Fig \ref{fig:H_E-static-1}, we see that the expectation values in the state with $t = 1/2j_0$ follow the (constant) semiclassical trajectory extremely closely throughout the time interval over which the computation is performed. The occupation number $n_{\rm max}$ remains small over most of the time interval, but starts somewhat increasing near the end, suggesting that the approximation is beginning to break down. The state remains well peaked on the observables through most of its evolution, although there is a moderate amount of spreading with respect to the volume operator.

In contrast, \Fig \ref{fig:H_E-static-2} exhibits a very different behavior for the state with $t = 1/2\sqrt{j_0}$. The state is initially much more sharply peaked on the volume operator, with a correspondingly wider spread with respect to the holonomy operators. As the state evolves, the dispersion increases rather rapidly, which causes the cutoff to be reached at a much earlier stage of the computation. From time $T \simeq 20$ onwards the occupation number $n_{\rm max}$ has grown so large that the approximation should probably no longer be considered reliable. Therefore we are unable to conclusively say whether the observed departure of the expectation values from the semiclassical trajectory, which indicates the state having lost its semiclassical properties, is a true feature of the quantum dynamics or merely an artifact caused by the cutoff.

Similar behavior can be observed for the states in which $t = 1/2\sqrt{j_0}$ also in other examples, with initial data corresponding to a non-static geometry. For this reason we settled on the choice $t = 1/2j_0$, as opposed to $t \sim 1/\sqrt{j_0}$, as a value of the semiclassicality parameter that seems to yield coherent states displaying better semiclassical properties under the dynamics.

\subsubsection*{Expanding universe}

\begin{figure}[p]
	\centering
	\includegraphics[width=\textwidth]{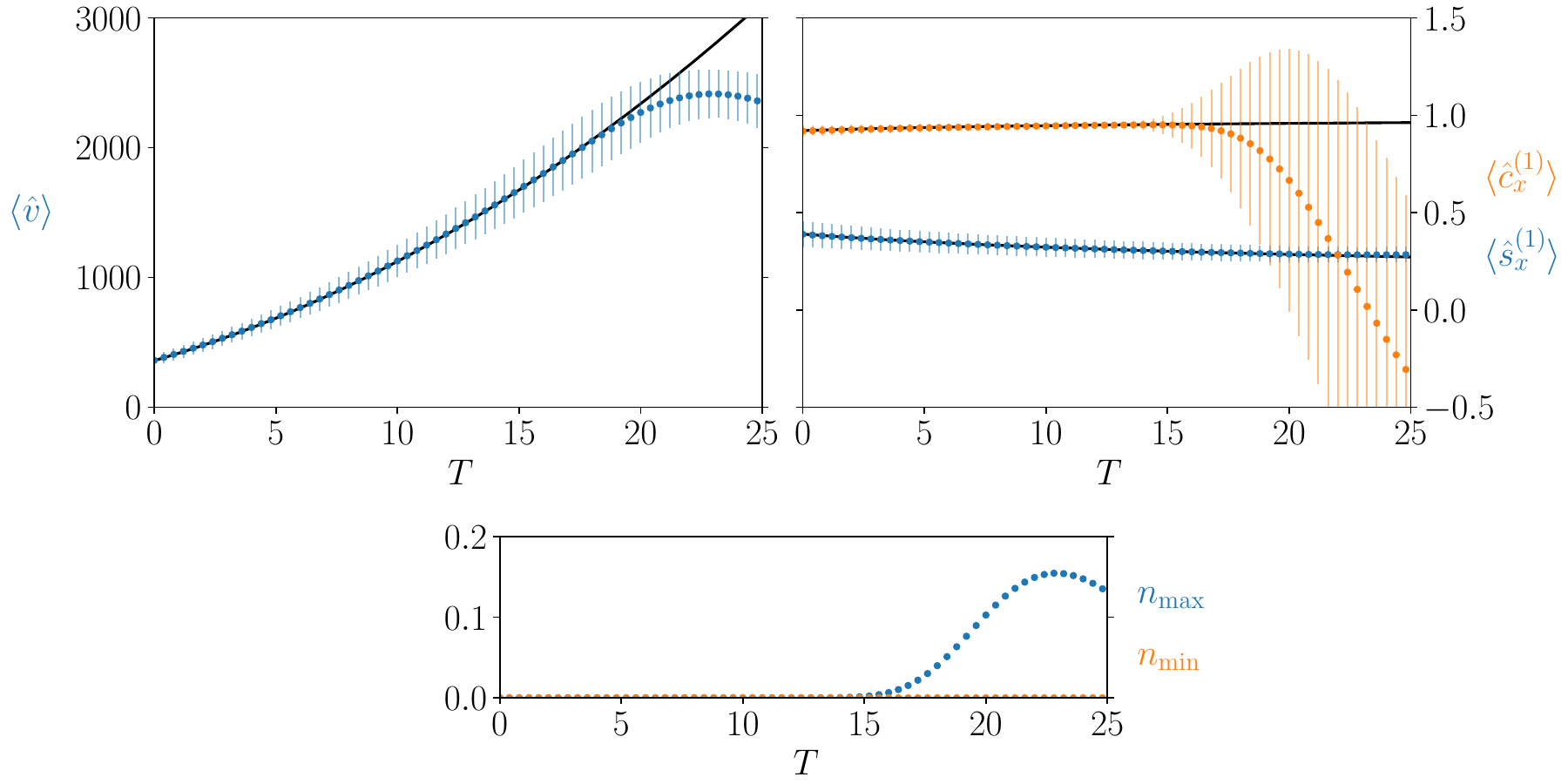}
	\caption{Time evolution of a state describing an expanding universe in the Euclidean model. The initial state is defined by the parameters $j_0 = 50$, $c_0 = 0.4$. The quantum dynamics matches the semiclassical trajectory and the state remains well peaked throughout the time interval over which the computation can be expected to be accurate.}
	\label{fig:H_E-expanding-1}
\end{figure}

\begin{figure}[p]
	\centering
	\includegraphics[width=\textwidth]{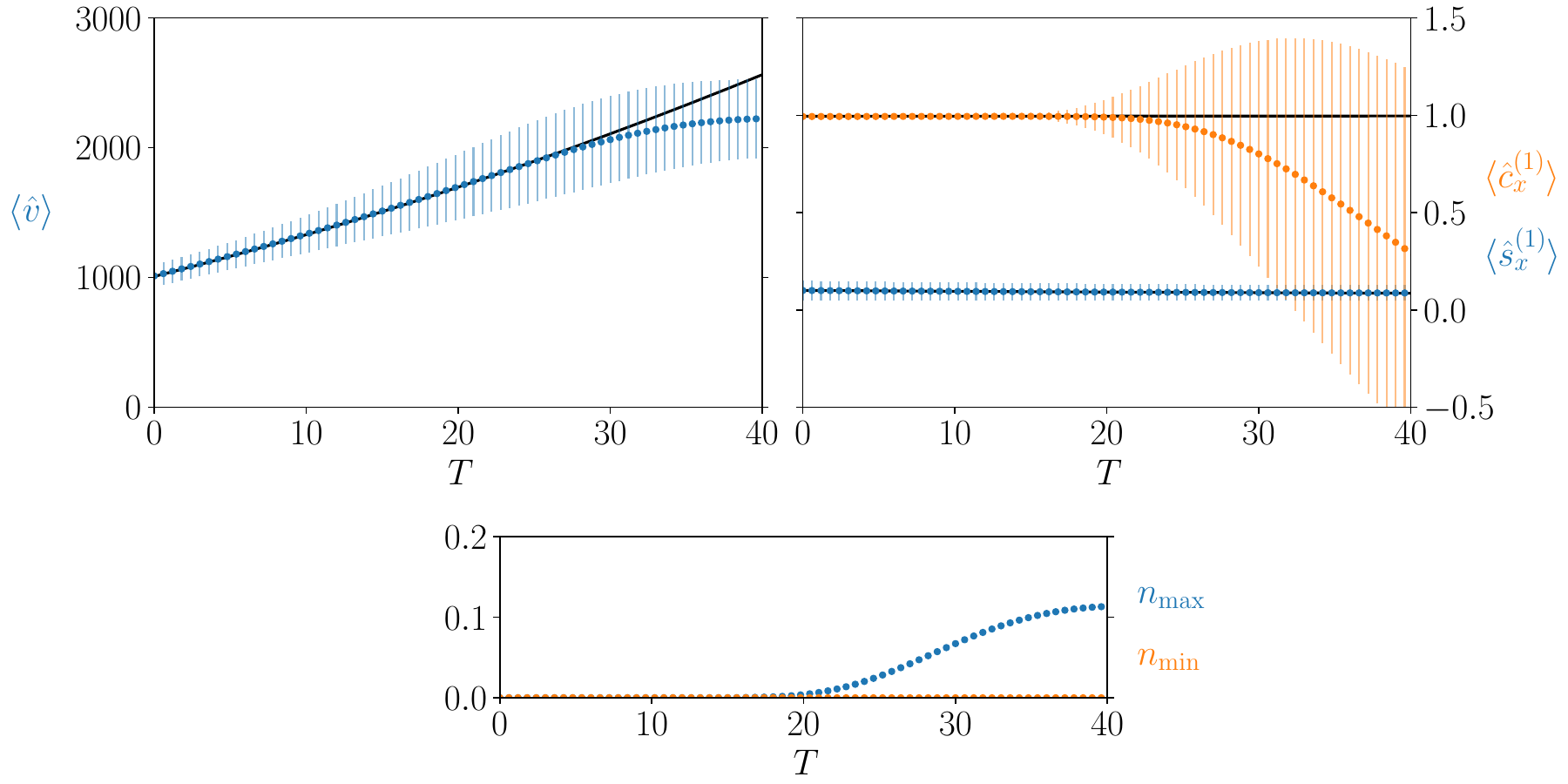}
	\caption{Another example of a state describing an expanding universe in the Euclidean model. Here the parameters of the initial state are $j_0 = 100$, $c_0 = 0.1$. The qualitative features of the dynamics are essentially identical to the example shown in \Fig \ref{fig:H_E-expanding-1}.}
	\label{fig:H_E-expanding-2}
\end{figure}

\Figs \ref{fig:H_E-expanding-1} and \ref{fig:H_E-expanding-2} present two examples in which the parameter $c_0$ is set to a positive value, and which therefore describe an expanding universe. The initial states are defined respectively by the values $j_0 = 50$ and $c_0 = 0.4$ in \Fig \ref{fig:H_E-expanding-1}, and $j_0 = 100$ and $c_0 = 0.1$ in \Fig \ref{fig:H_E-expanding-2}. Thus, the first state begins at a lower value of the volume and expands at a faster rate, while the second state, being characterized by a higher value of $j_0$ and a lower value of $c_0$, has larger initial volume and a slower rate of expansion. Both examples display a very similar qualitative behavior. The expectation values of all the observables are very closely centered on the effective semiclassical trajectory, and the state remains well peaked on them, until the point at which the occupation number $n_{\rm max}$ has attained an appreciably large value, signifying that the computation has exceeded its range of validity. Qualitatively, one can say that the time scale after which the cutoff begins interfering with the computation is determined by some combination of the rate of expansion and the rate of spreading of the wave packet.

\begin{figure}[p]
	\centering
	\includegraphics[width=\textwidth]{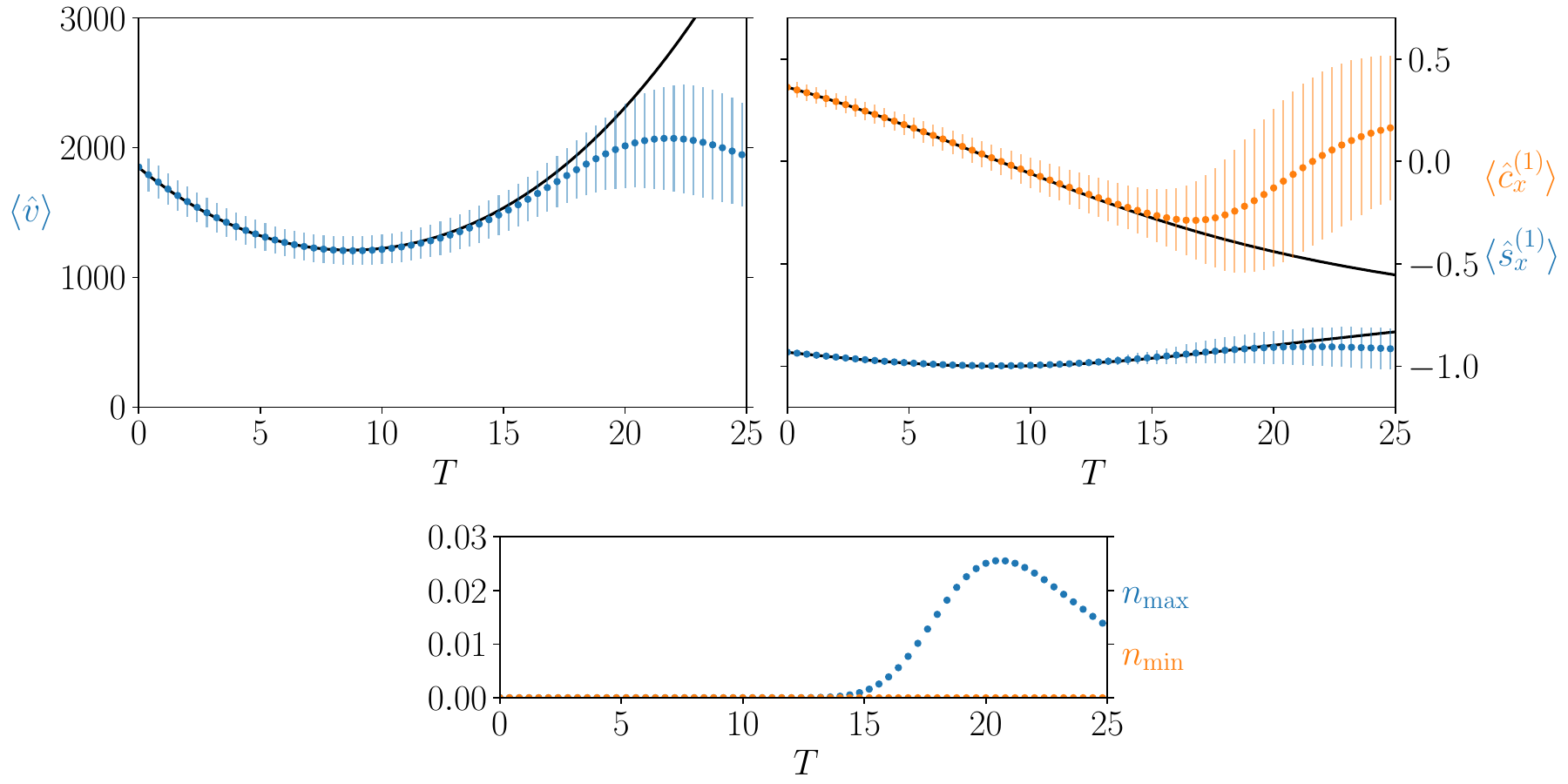}
	\caption{An example of an initially contracting state undergoing a dynamical ``bounce'' in the Euclidean model. The initial state is described by $j_0 = 150$, $c_0 = -1.2$. The state remains sharply peaked and the expectation values follow the semiclassical trajectory through the bounce, beginning to deviate only around the time the occupation number $n_{\rm max}$ has grown to a significant value.}
	\label{fig:H_E-bounce-1}
\end{figure}

\begin{figure}[p]
	\centering
	\includegraphics[width=\textwidth]{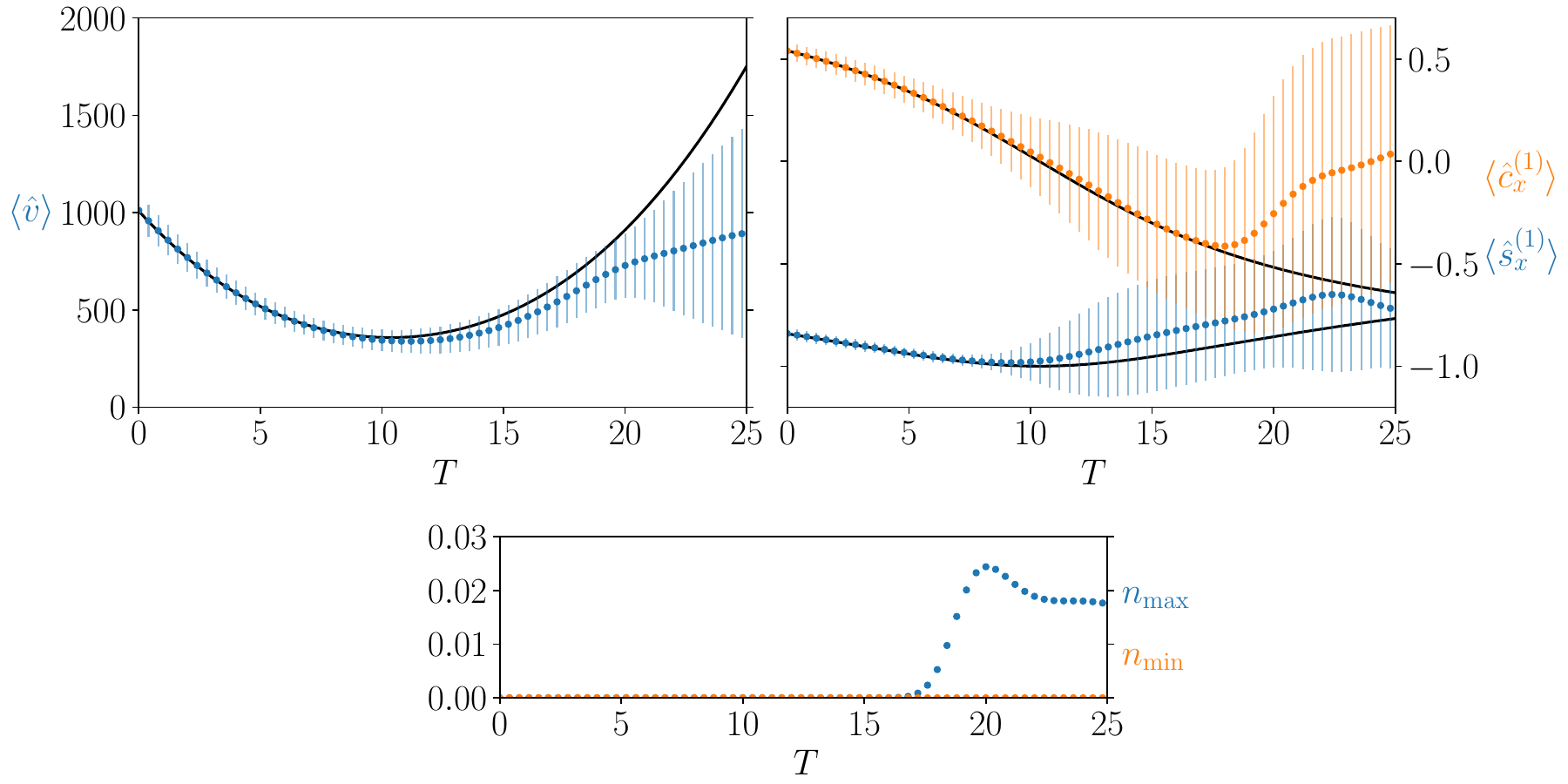}
	\caption{Another example of a state undergoing a ``quantum bounce'' in the Euclidean model. The parameters of the initial state are given by $j_0 = 100$, $c_0 = -1$. Here the expectation values start to deviate noticeably from the semiclassical trajectory already around the bounce, while the occupation number $n_{\rm max}$ is still negligibly small. Shortly after the bounce, the state no longer seems to be well peaked on the holonomy operators.}
	\label{fig:H_E-bounce-2}
\end{figure}

\subsubsection*{Contracting universe}

In \Figs \ref{fig:H_E-bounce-1} and \ref{fig:H_E-bounce-2} we consider two examples of states describing an initially contracting universe, characterized by a negative value of the parameter $c_0$. Classically, the dynamics of a homogeneous and isotropic universe with a negative initial rate of expansion terminates in a singularity after a finite time. In the context of loop quantum cosmology and related models, it has been conclusively established that the classical singularity is resolved in the quantum theory: An initially contracting quantum state dynamically undergoes a ``bounce'' and enters an expanding phase (see \eg\cite{Li:2023dwy, Gambini:2022hxr} and references therein). This behavior is also borne out at the semiclassical level by the dynamics generated by the effective Hamiltonian \eqref{eq:H_eff^E}.

The results shown in \Figs \ref{fig:H_E-bounce-1} and \ref{fig:H_E-bounce-2} confirm that a quantum bounce is also exhibited by the dynamics of the one-vertex model considered here: The initially contracting evolution of the state is halted and turned into an expansion by the quantum dynamics. At the time at which the bounce occurs, both occupation numbers $n_{\rm max}$ and $n_{\rm min}$ are vanishingly small. The former indicates that the numerical computation is still providing an accurate approximation of the dynamics and the results are not yet being affected by the presence of the cutoff, while the latter suggests that the bounce is a genuine feature of the quantum dynamics, instead of being some kind of an artifact caused by the lower limit $j_{\rm min} = 1$ (which, while not being an artificial, unphysical cutoff in the same way as the upper cutoff at $j = j_{\rm max}$, is still a consequence of a specific choice of factor ordering for the Hamiltonian, and would certainly not be present for every possible choice of factor ordering). 

In terms of comparing the evolution of the quantum observables with the effective semiclassical dynamics, we have observed two qualitatively different kinds of behavior. \Fig \ref{fig:H_E-bounce-1}, in which the initial state is labeled by $j_0 = 150$ and $c_0 = -1.2$, provides an example where the quantum dynamics of the state matches the semiclassical trajectory very accurately throughout the bounce; moreover, the state remains well peaked on the observables and starts visibly losing its semiclassical properties only after the time at which the quantity $n_{\rm max}$ has grown significantly large.

A markedly different type of behavior is illustrated in \Fig \ref{fig:H_E-bounce-2}, where the parameters of the initial state are given by $j_0 = 100$ and $c_0 = -1$. Here the expectation values of the volume operator $\hat v$ and the holonomy operator $\hat s_x^{(1)}$ begin to deviate noticeably from the semiclassical trajectory already in the region around the bounce. Moreover, the dispersion of both holonomy operators rapidly increases, and the state at a time shortly after the bounce no longer appears to be a good semiclassical state in terms of its kinematical peakedness properties. As pointed out above, the occupation number $n_{\rm max}$ is still very small in the region around the bounce, so the observed behavior presumably is an accurate description the true dynamics of the state.

In general it seems that the type of behavior seen in \Fig \ref{fig:H_E-bounce-1} occurs more often in states where the bounce takes place at a higher value of the volume, whereas the behavior in \Fig \ref{fig:H_E-bounce-2} is more typical to states for which the value of the volume at the bounce is lower. A plausible preliminary explanation of this behavior would therefore be connected to the fact that in quantum-reduced loop gravity the spins are conventionally required to be large, as discussed in section \ref{sec:qrlg}, and it is only in the regime of large spins that the reduced operators are closely related to the operators of full loop quantum gravity in the way indicated by \Eqs \eqref{eq:O}--\eqref{eq:O^R}. Thus, it seems possible that the smallest spins of the basis states involved in the dynamics of the state in \Fig \ref{fig:H_E-bounce-2} are not large enough to be considered ``large'' in the sense of a large-$j$ approximation.

\subsection{Lorentzian model}
\label{sec:H_L}

From the perspective of the numerical methods used to perform our calculations, the Lorentzian Hamiltonian \eqref{eq:H_phys} seems to present a problem of significantly higher computational complexity than the Euclidean Hamiltonian $\hat H_E$ alone. The Python program, which was used successfully in the Euclidean model alongside the Julia program, would require an impractically long time to compute the time evolution of a given initial state under the full Lorentzian Hamiltonian for the chosen cutoff value $j_{\rm max} = 200$. For computations in the Lorentzian model, we are therefore restricted to using the Julia program only. In an attempt to confirm the reliability of the results in this case, the computation for a given initial state is performed twice, using two different values for the dimension of the so-called Krylov subspace, which is a parameter that controls the accuracy of the calculation and which can be passed an argument to the function \texttt{expv}. When both computations produce essentially identical results, we take this as a reasonable indication that the results are indeed accurate.

Due to the increased numerical complexity, and accordingly longer computation times required for the Lorentzian model, the range of examples we have been able to analyze is not as extensive as in the case of the Euclidean model. Nevertheless, we have computed the dynamics of several different states describing expanding and contracting geometries, in addition to the example of a static universe. In all calculations performed in the Lorentzian model, the value of the Barbero--Immirzi parameter is set to $\beta = 1$.

\subsubsection*{Static universe}

\begin{figure}[p]
	\centering
	\includegraphics[width=\textwidth]{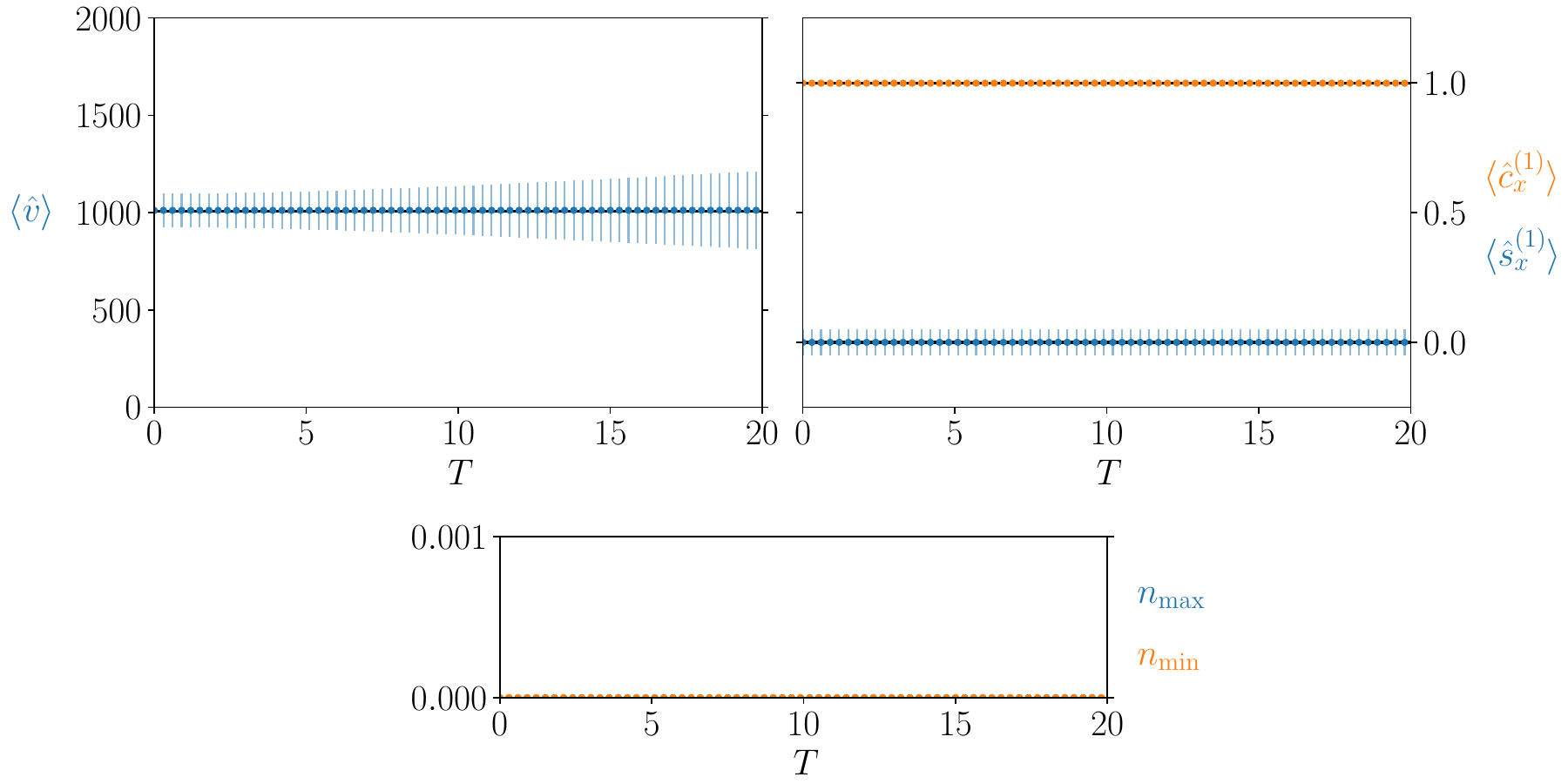}
	\caption{Evolution of a state peaked on data describing a static universe in the Lorentzian model. The parameters of the initial state are given by $j_0 = 100$, $c_0 = 0$. The quantum dynamics of the state matches the effective semiclassical trajectory very accurately. Over the time interval considered, only a moderate spread can be seen in the dispersion for the volume operator.}
	\label{fig:H_L-static}
\end{figure}

The time evolution of an initial state corresponding to a classically static configuration is displayed in \Fig \ref{fig:H_L-static}. The initial state is again described by the parameters $j_0 = 100$ and $c_0 = 0$. The expectation values of the observables in the evolving state again match the effective semiclassical trajectory very sharply throughout its evolution. Over the time interval included in the computation, only a moderate amount of spreading can be seen in the dispersion of the volume operator, while the dispersions of the holonomy operators remain virtually unchanging. Due to the long computation time required to carry out the calculation over longer time intervals, the evolution in this example was computed only until $T = 20$, instead of pushing it all the way to the end of its range of validity (\ie the point at which the value of $n_{\rm max}$ has grown to be non-negligible).

\subsubsection*{Expanding universe}

\begin{figure}[p]
	\centering
	\includegraphics[width=\textwidth]{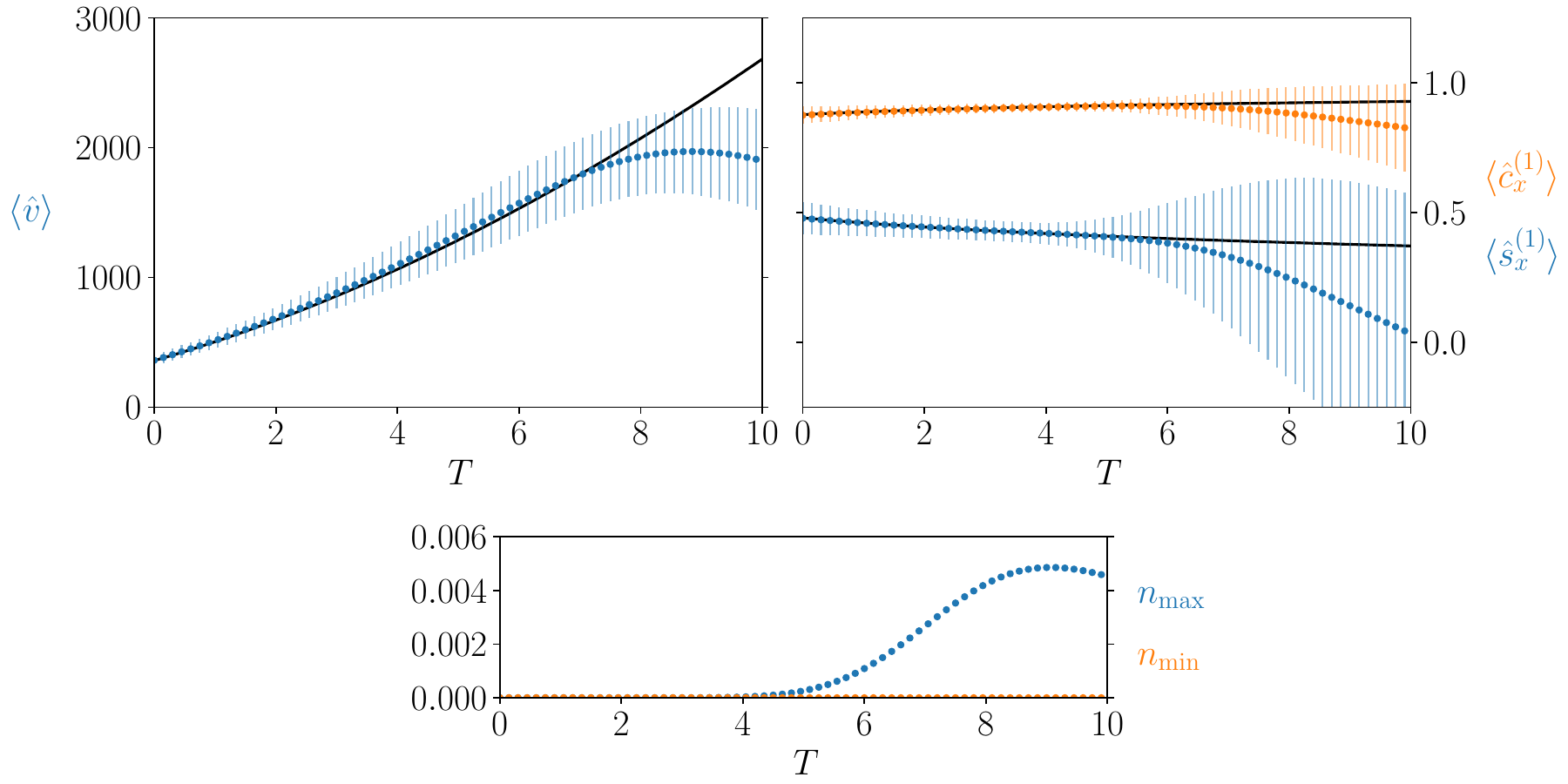}
	\caption{Dynamics of a state corresponding to an expanding universe in the Lorentzian model. The parameters describing the initial state are $j_0 = 50$, $c_0 = 0.5$. The evolution of the expectation values again closely follows the semiclassical effective dynamics, until the point at which the presence of the cutoff begins affecting the computation.}
	\label{fig:H_L-expanding}
\end{figure}

\Fig \ref{fig:H_L-expanding} presents an example of the dynamics of a state describing an expanding universe in the Lorentzian model. The parameters of the initial state are given by $j_0 = 50$ and $c_0 = 0.5$. The qualitative behavior is again very similar to the analogous examples seen in the Euclidean model. The evolution of the quantum state closely follows the semiclassical trajectory throughout the time interval over which the numerical computation can be expected to accurately describe the dynamics, until the presence of the cutoff begins to disrupt the evolution, as indicated by the increasing occupation number $n_{\rm max}$.

\begin{figure}[p]
	\centering
	\includegraphics[width=\textwidth]{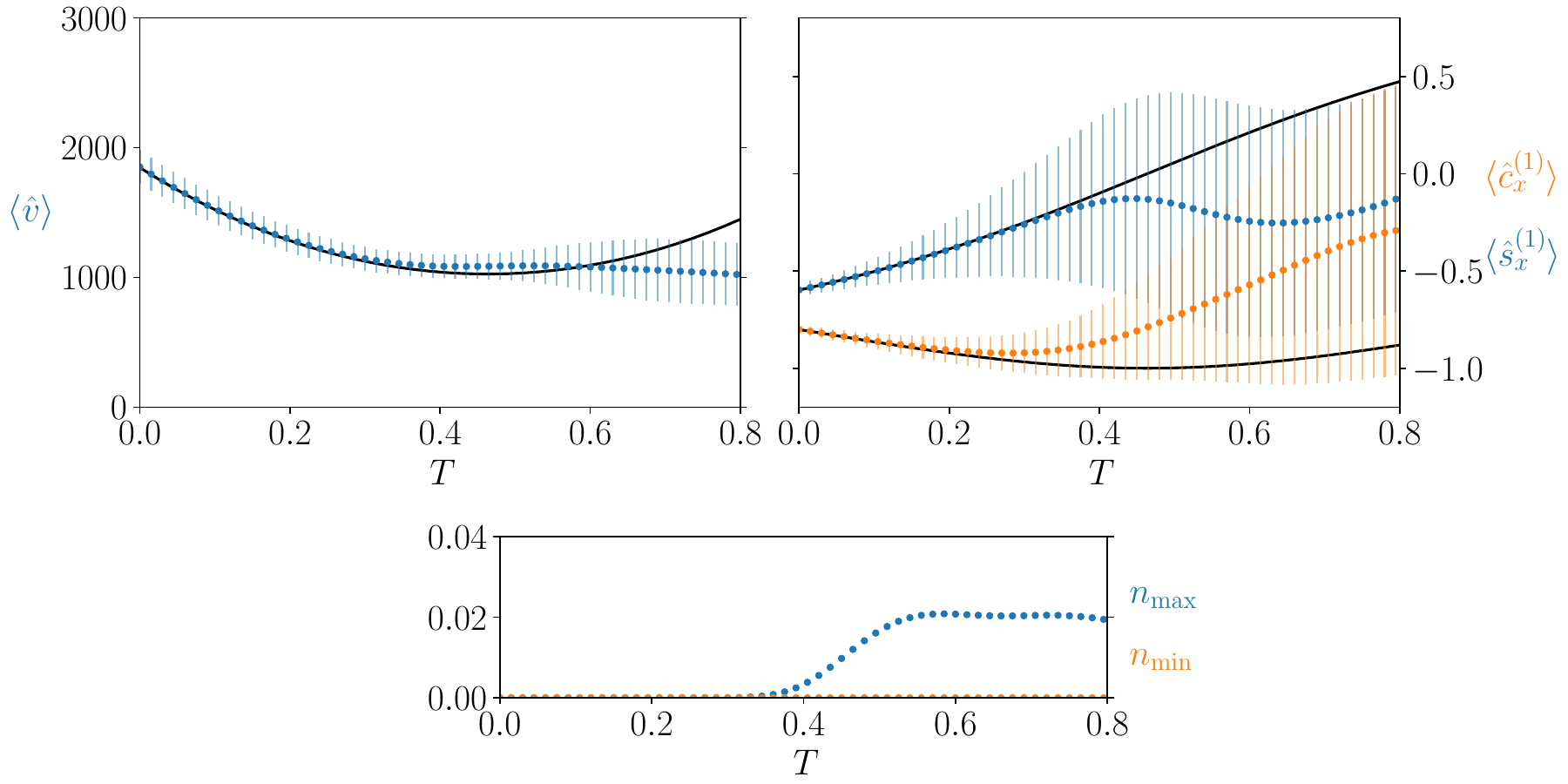}
	\caption{An example of the evolution of an initially contracting state in the Lorentzian model. The initial state is labeled by the parameters $j_0 = 150$, $c_0 = -2.5$. The occupation number $n_{\rm max}$ grows to a substantial value already before the bounce has occurred, implying that the computation is not providing an accurate description of the dynamics at or after the bounce.}
	\label{fig:H_L-bounce-1}
\end{figure}

\begin{figure}[p]
	\centering
	\includegraphics[width=\textwidth]{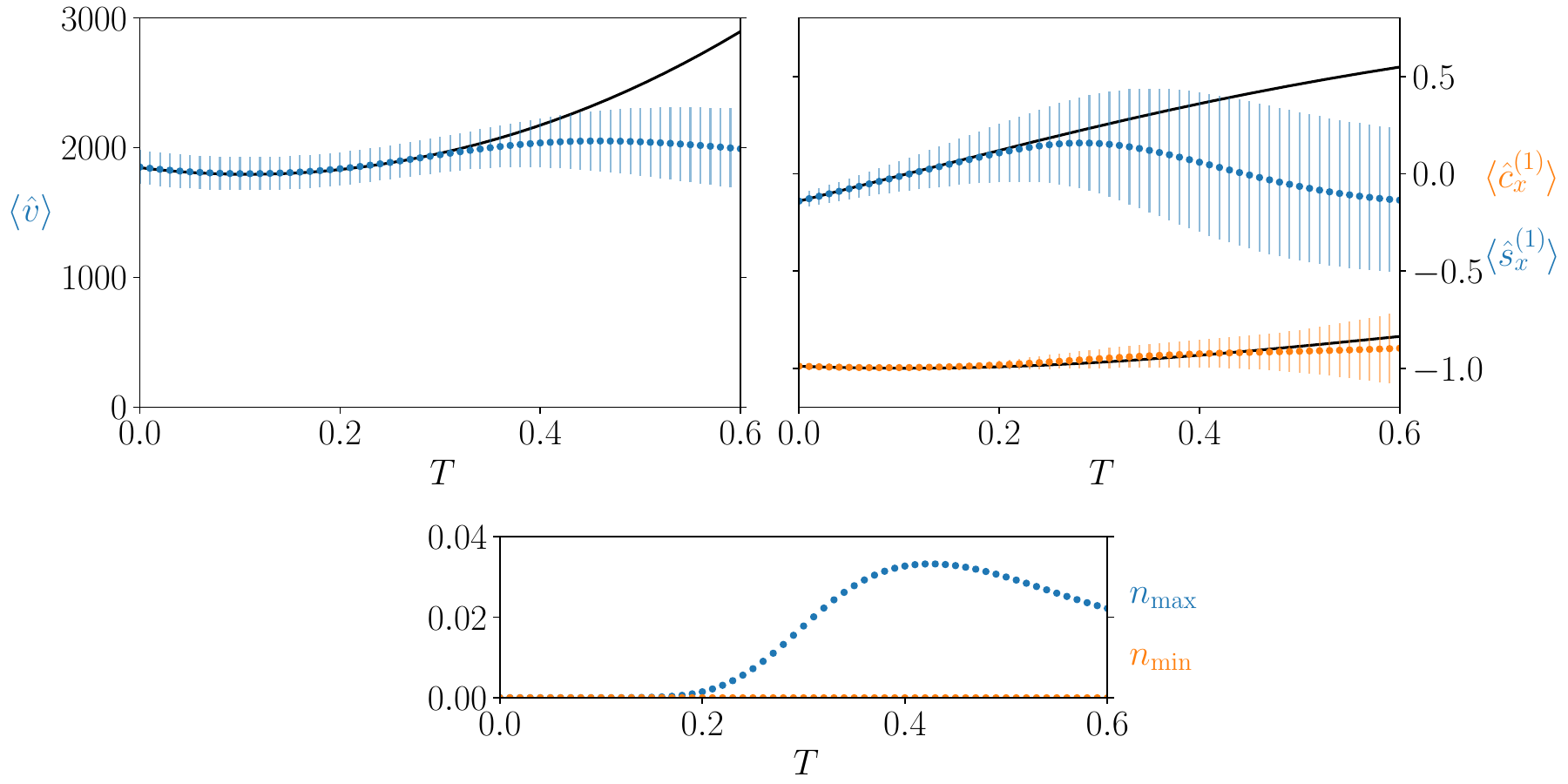}
	\caption{Evolution of an initial state in the Lorentzian model which is peaked on a point of the effective semiclassical trajectory immediately before the bounce. The parameters defining the state are $j_0 = 150$, $c_0 = -3$. Here the state evolves through the bounce and undergoes a brief period of expansion while remaining semiclassical, before the computation is disrupted by the cutoff.}
	\label{fig:H_L-bounce-2}
\end{figure}

\subsubsection*{Contracting universe}

When attempting to study examples of states going through a quantum bounce in the Lorentzian model, we encounter the following problem: As a state describing an initially contracting geometry evolves, it seems that the basis states adjacent to the cutoff become occupied much more rapidly than in the Euclidean model. This makes it difficult to prepare an initial state that lies some distance away from the bounce on the effective semiclassical trajectory, while being able to evolve the state through the bounce without the computation becoming unreliable due to the cutoff having been reached. An example is given in \Fig \ref{fig:H_L-bounce-1}, where the initially well-peaked state clearly loses its semiclassical properties before it has completed the bounce, and based on the value of the occupation number $n_{\rm max}$ it seems reasonable to attribute this to the numerical approximation having broken down due to the presence of the cutoff. The initial state in this example is described by the parameters $j_0 = 150$ and $c_0 = -2.5$.

The closest we have been able to come to producing a satisfactory example of a bounce in the Lorentzian model is to choose the parameters of the initial state such that the state lies on a point of the semiclassical trajectory very shortly before the bounce. An initial state of this type can evolve through the bounce while following the semiclassical trajectory and remaining well peaked on the volume and holonomy operators, and emerge (barely) into an expanding phase before the occupation number $n_{\rm max}$ begins to grow large and the evolution of the state is disrupted. An example is provided in \Fig \ref{fig:H_L-bounce-2}, where the parameters labeling the initial state are given by $j_0 = 150$ and $c_0 = -3$. This makes it seem plausible that if we were able to perform computations for the Lorentzian model at a substantially higher value of the cutoff, we should be able to find examples analogous to that shown in \Fig \ref{fig:H_E-bounce-1} in the Euclidean model, where the state undergoes a bounce while retaining its semiclassical properties.

\section{Remark on a parametrized family of Hamiltonians}
\label{sec:H_abc}

During the course of investigating the behavior observed in certain examples in the Euclidean model, in which the quantum dynamics of an initially contracting state begins to deviate from the effective semiclassical dynamics at an unexpectedly early stage of the evolution, we made an observation which we believe is worth being reported here, even though its meaning or proper interpretation is perhaps not very clear at the moment. The observation stems from studying the quantum dynamics generated by a parametrized family of Hamiltonian operators in the one-vertex model. These operators are defined as
\begin{equation}
	\hat H_{\rm I}^{(\alpha\beta\gamma)} = -\frac{1}{2}\biggl(\bigl(\hat p_x^\alpha\hat p_y^\beta\hat p_z^\gamma\bigr)^{1/2}\hat s_x^{(1)}\hat s_y^{(1)}\bigl(\hat p_x^\alpha\hat p_y^\beta\hat p_z^\gamma\bigr)^{1/2} + \text{permutations} \biggr)
	\label{eq:H_I}
\end{equation}
and
\begin{equation}
	\hat H_{\rm II}^{(\alpha\beta\gamma)} = -\frac{1}{2}\biggl(\bigl(\hat p_x^\alpha\hat p_y^\beta\hat p_z^\gamma\bigr)^{1/2}\bigl(\hat s_x^{(1)}\bigr)^2\bigl(\hat p_x^\alpha\hat p_y^\beta\hat p_z^\gamma\bigr)^{1/2} + \text{permutations} \biggr).
	\label{eq:H_II}
\end{equation}
Here ``permutations'' denotes the terms obtained from the first term by making all permutations of the labels $x$, $y$ and $z$, so the resulting operator is symmetric with respect to the three coordinate directions.

For generic values of the parameters $\alpha$, $\beta$ and $\gamma$, the operators \eqref{eq:H_I} and \eqref{eq:H_II} of course do not have any immediate physical significance\footnote{
	Besides the operator \eqref{eq:H_I} with parameters $\alpha = \beta = 1/2$ and $\gamma = -1/2$, which corresponds to the Euclidean operator $\hat H_E$ given by \Eq \eqref{eq:H_E}, another case which can be assigned a direct physical interpretation is the operator \eqref{eq:H_II} with $\alpha = 1/2$ and $\beta = \gamma = 0$, which is the sum of three independent, one-dimensional operators formally identical to the Hamiltonian constraint of homogeneous and isotropic loop quantum cosmology in the $\mu_0$ scheme.
}.
In particular, they should not be thought of as representing valid, physically correct quantizations of the Euclidean physical Hamiltonian for the one-vertex model. Their role is rather to serve as a family of mathematical toy examples, which can provide some insight as to which features of the Hamiltonian in the Euclidean model might be contributing to the seemingly peculiar behavior of some of the bouncing examples. The motivation to study these examples can be explained by considering the effective dynamics generated by the corresponding effective Hamiltonians
\begin{equation}
	H_{\rm I}^{(\alpha\beta\gamma)} = -\frac{1}{2}\biggl(p_x^\alpha p_y^\beta p_z^\gamma\frac{\sin\mu c_x}{\mu}\frac{\sin\mu c_y}{\mu} + \text{permutations}\biggr)
	\label{eq:H^eff_I}
\end{equation}
and
\begin{equation}
	H_{\rm II}^{(\alpha\beta\gamma)} = -\frac{1}{2}\biggl(p_x^\alpha p_y^\beta p_z^\gamma\frac{\sin^2\mu c_x}{\mu^2} + \text{permutations}\biggr)
	\label{eq:H^eff_II}
\end{equation}
on the classical phase space of a spatially homogeneous but anisotropic universe, which is coordinatized by the three connections $c_a$ and the canonically conjugate triad variables $p_a$:
\begin{equation}
	\{c_a, p_b\} \sim \delta_{ab}.
	\label{}
\end{equation}
Provided that the parameters $\alpha$, $\beta$ and $\gamma$ are constrained by the condition
\begin{equation}
	\alpha + \beta + \gamma = \frac{1}{2}
	\label{}
\end{equation}
all of the Hamiltonians \eqref{eq:H^eff_I} and \eqref{eq:H^eff_II} generate the same effective dynamics for isotropic initial data (\ie $c_x = c_y = c_z$ and $p_x = p_y = p_z$ at time $T = 0$). Hence we may study the quantum dynamics of a given initial state under the different Hamiltonians \eqref{eq:H_I} and \eqref{eq:H_II} and examine any potential differences in how well the evolution matches the effective semiclassical dynamics common to all the Hamiltonians.

\begin{figure}[p]
	\centering
	\includegraphics[width=\textwidth]{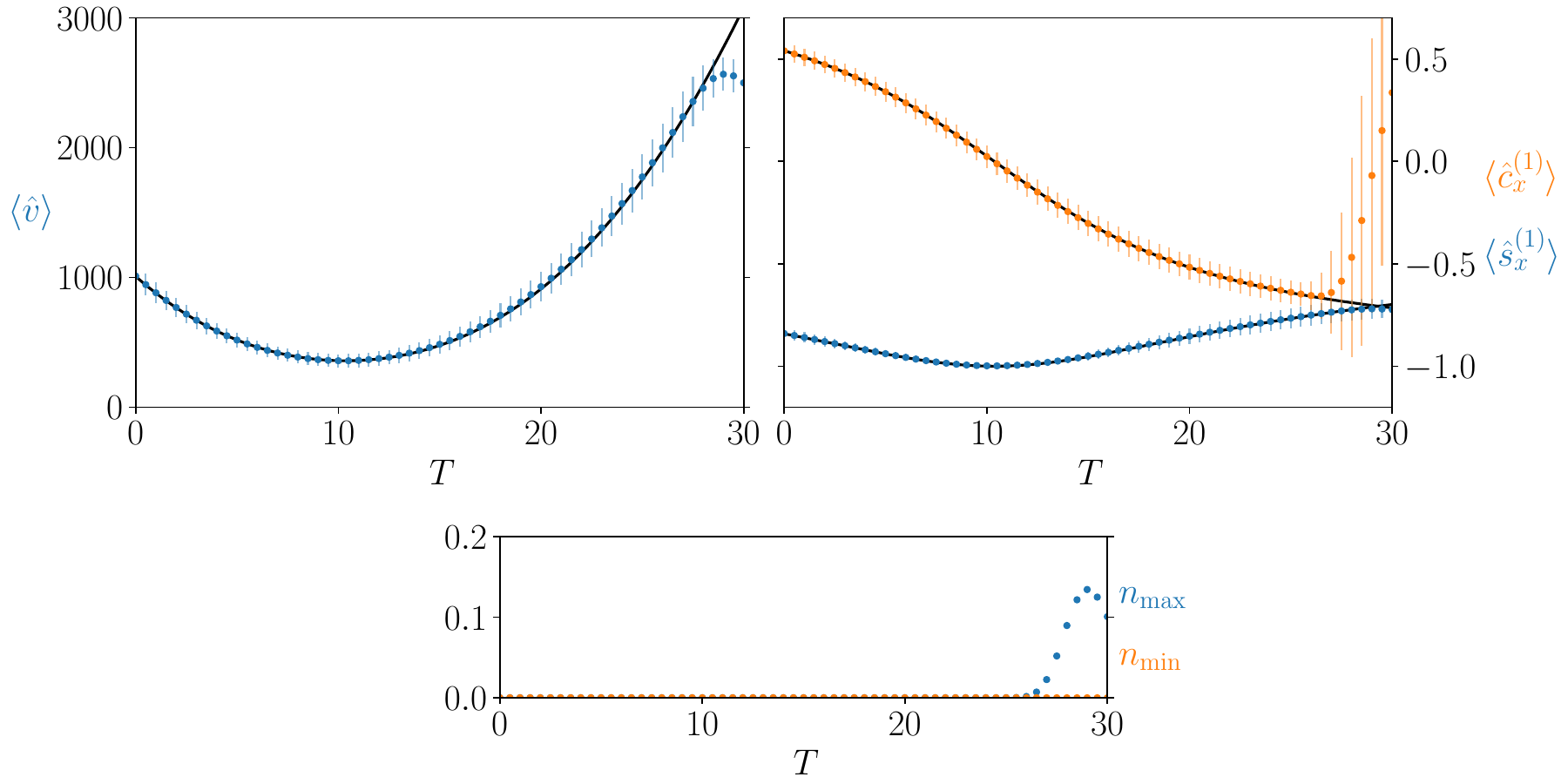}
	\caption{Evolution of the initial state described by the parameters $j_0 = 100$ and $c_0 = -1$ under the Hamiltonian $\hat H_{\rm I}^{(\alpha\beta\gamma)}$ with $\alpha = \beta = 0$ and $\gamma = 1/2$. While the cutoff is not reached, the quantum dynamics matches the effective dynamics very closely, and the state remains well peaked on the semiclassical trajectory.}
	\label{fig:H_abc-good-1}
\end{figure}

\begin{figure}[p]
	\centering
	\includegraphics[width=\textwidth]{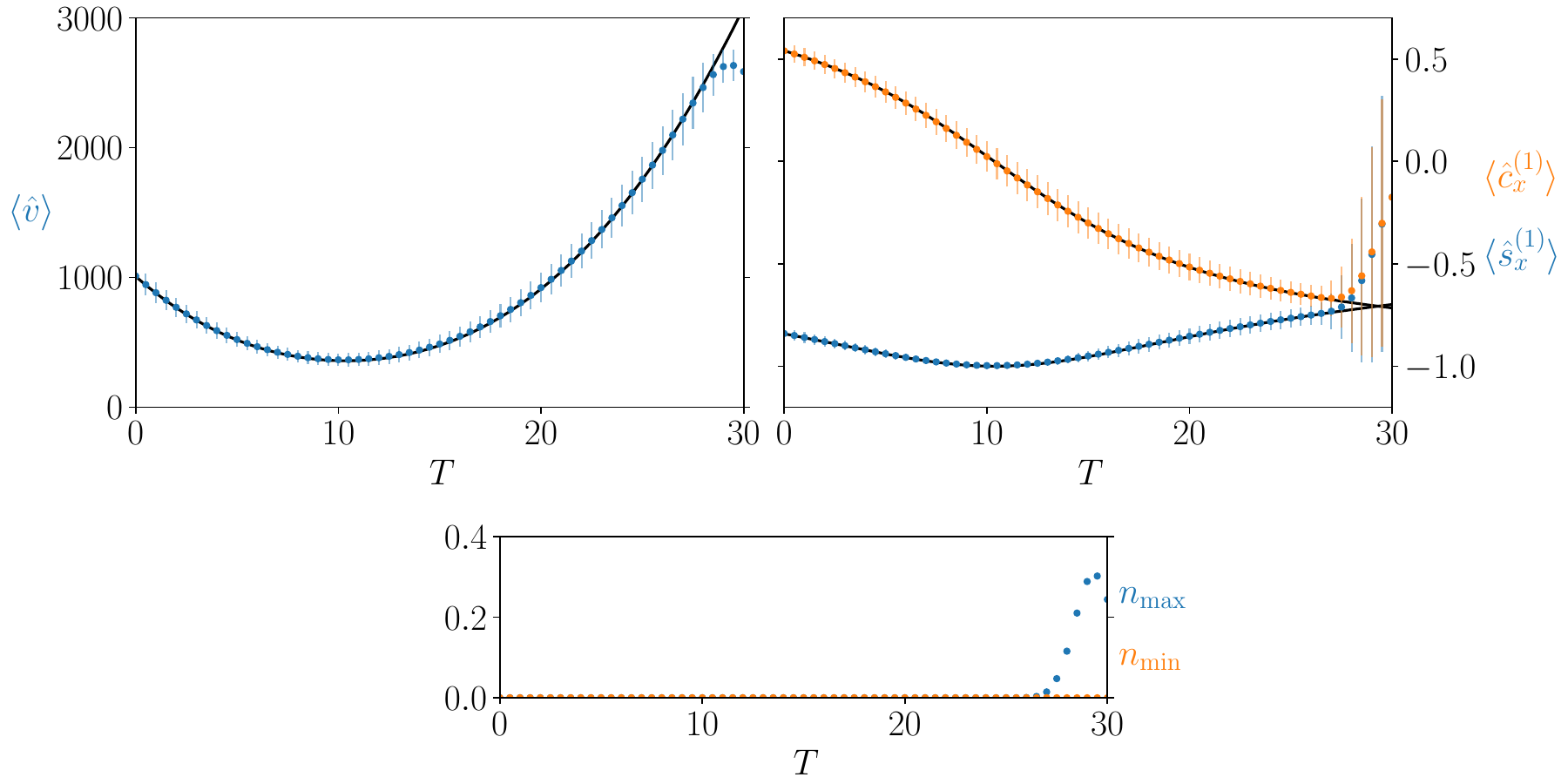}
	\caption{Evolution of the same initial state generated by the Hamiltonian $\hat H_{\rm II}^{(\alpha\beta\gamma)}$ with $\alpha = \beta = \gamma = 1/6$. Over the time interval where the cutoff is not yet relevant, the dynamics of the state seems to be virtually identical under any of the Hamiltonians \eqref{eq:H_I} and \eqref{eq:H_II}, as long as all the parameters $\alpha$, $\beta$ and $\gamma$ are non-negative.}
	\label{fig:H_abc-good-2}
\end{figure}

\begin{figure}[p]
	\centering
	\includegraphics[width=\textwidth]{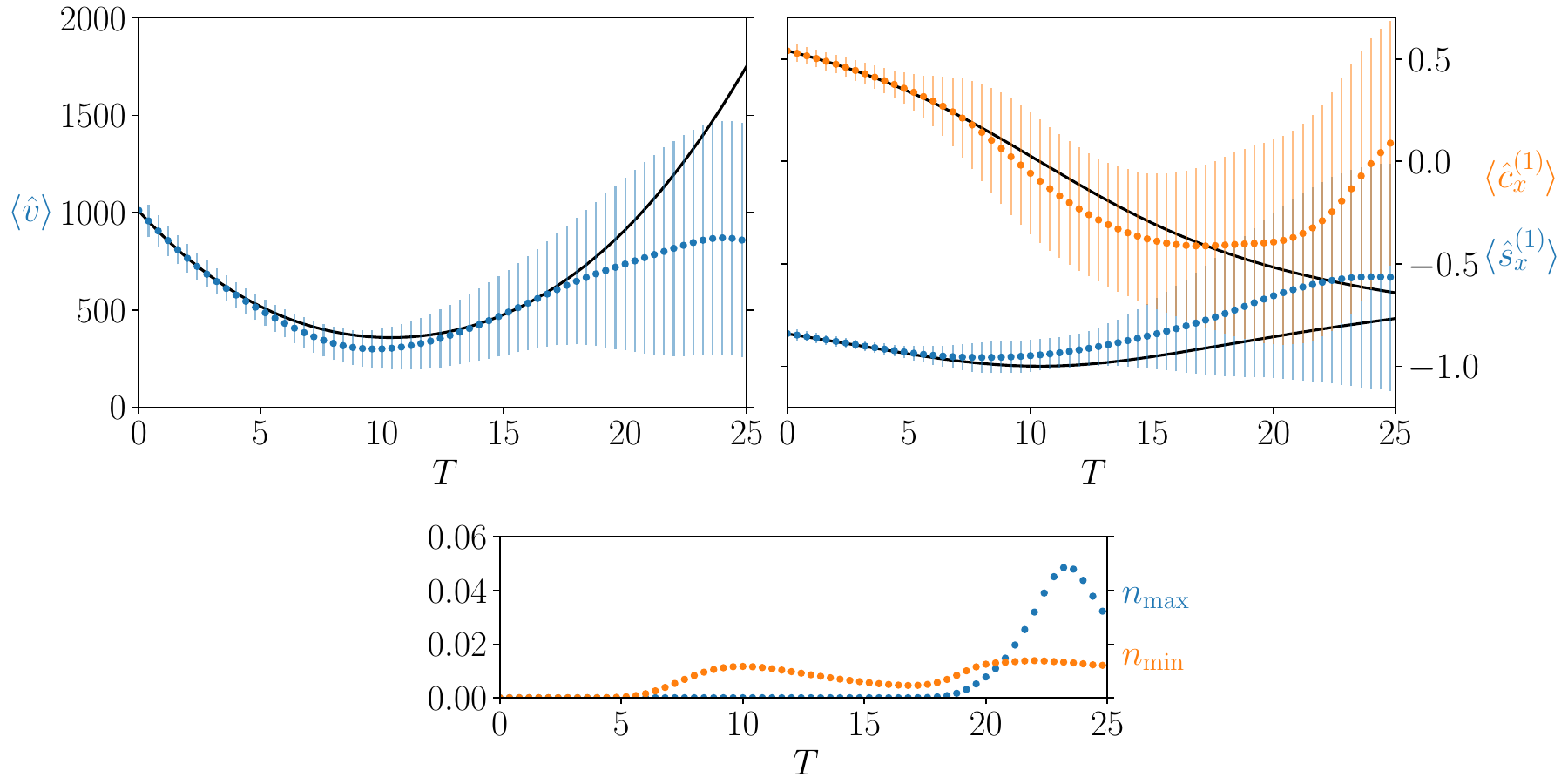}
	\caption{Dynamics of the initial state defined by $j_0 = 100$ and $c_0 = -1$ under the Hamiltonian $\hat H_{\rm I}^{(\alpha\beta\gamma)}$ with $\alpha = \beta = -1/4$ and $\gamma = 1$. The quantum dynamics begins to diverge from the effective semiclassical dynamics already at an early stage of the evolution.}
	\label{fig:H_abc-bad-1}
\end{figure}

\begin{figure}[p]
	\centering
	\includegraphics[width=\textwidth]{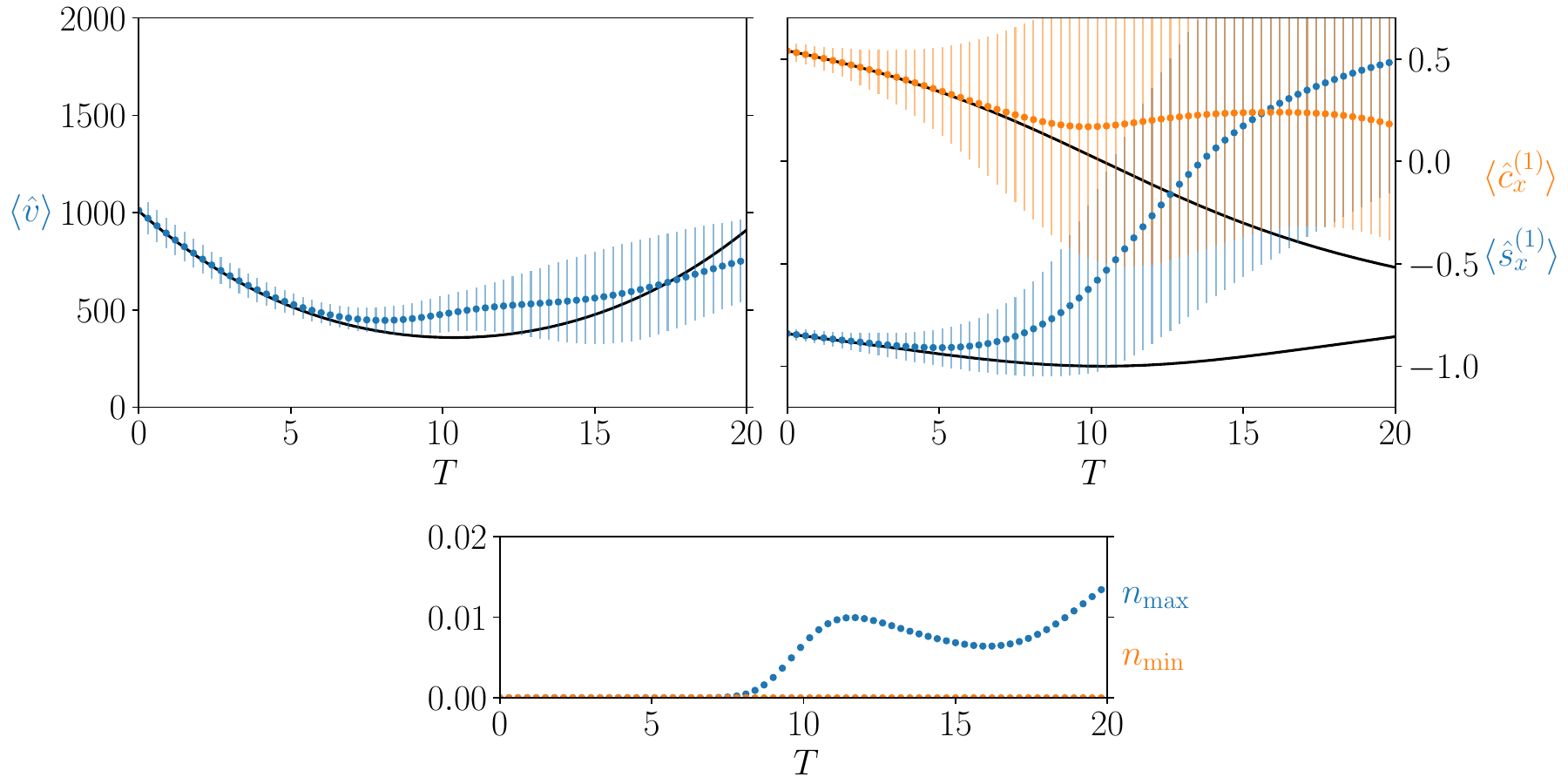}
	\caption{Dynamics of the same initial state under the Hamiltonian $\hat H_{\rm II}^{(\alpha\beta\gamma)}$ with the parameters given by $\alpha = 3/2$ and $\beta = \gamma = -1/2$. In the examples in which some of the parameters labeling the Hamiltonian are negative, the evolution of the quantum state consistently fails to follow the semiclassical trajectory through the bounce.}
	\label{fig:H_abc-bad-2}
\end{figure}

To this end, we computed the time evolution of the state $\ket{\psi_{j_0,c_0}}$ with $j_0 = 100$ and $c_0 = -1$ generated by the operators \eqref{eq:H_I} and \eqref{eq:H_II} for a selection of different values of the parameters $\alpha$, $\beta$ and $\gamma$. This is the initial state whose dynamics under the Euclidean Hamiltonian $\hat H_E$ is displayed in \Fig \ref{fig:H_E-bounce-2}. From the results of the calculations, a clear pattern seems to emerge. If all the parameters $\alpha$, $\beta$ and $\gamma$ are non-negative, or if one of them has a small negative value, the quantum dynamics follows the effective semiclassical trajectory extremely closely, and deviates noticeably from it only after the range of validity of the computation has been exceeded (as determined by the rough criterion of looking at the occupation number $n_{\rm max}$). However, if any of the parameters takes a negative value not too close to zero, the quantum dynamics tends to exhibit the type of behavior encountered in \Fig \ref{fig:H_E-bounce-2}, where the evolution of the quantum expectation values begins to diverge from the semiclassical dynamics at a time when the occupation number $n_{\rm max}$ is still extremely small. Two examples of the former kind of behavior are given in \Figs \ref{fig:H_abc-good-1} and \ref{fig:H_abc-good-2}, while the latter kind of behavior is illustrated by the examples shown in \Figs \ref{fig:H_abc-bad-1} and \ref{fig:H_abc-bad-2} (in addition to the physically relevant example already seen in \Fig \ref{fig:H_E-bounce-2}).

It would of course be premature to draw any definite conclusions from the examples examined here, which all take place in the narrow setting of a specific type of initial state defined on the very simple one-vertex graph in quantum-reduced loop gravity. Nevertheless, I believe the results suggest that the semiclassical properties of the dynamics generated by a Hamiltonian constructed using the Tikhonov regularization of the inverse volume would be a question deserving of a more thorough investigation. (Recall that the Tikhonov regularization is the reason why factors of the operator $\hat p_a$ raised to the negative power $-1/2$ appear in the physical Hamiltonian discussed in section \ref{sec:H_phys}.) If the ``deteriorated'' dynamical behavior of semiclassical states under a Hamiltonian of this kind turns out to be a feature that can be reproduced consistently in a wider variety of examples, this could potentially be used as a concrete physical argument disfavoring the Tikhonov regularization and favoring alternative regularizations (in particular the Thiemann regularization) of the inverse volume element in the Hamiltonian. On the other hand, recalling the behavior seen in the example of \Fig \ref{fig:H_E-bounce-1} in contrast with \Fig \ref{fig:H_E-bounce-2}, it is also possible that our findings do not indicate any fundamental problem, and in order for good semiclassical dynamics to emerge from a Hamiltonian regularized using the Tikhonov prescription, it might simply be sufficient to consider states characterized by sufficiently high values of the spins.

\section{Conclusions}
\label{sec:conclusions}

In this article we examined the time evolution of semiclassical states representing homogeneous and isotropic spatial geometries in a simple model, which is obtained by considering states defined on a spin network graph formed by a single six-valent vertex in quantum-reduced loop gravity. The dynamics of the model is formulated in the deparametrized setting, with an irrotational dust field playing the role of a relational time variable for the dynamics of quantum states of the gravitational field. For the physical Hamiltonian operator, which defines the dynamics of the model, we use the operator studied previously in \cite{Makinen:2024rbg} in the context of quantum-reduced loop gravity and the one-vertex model; in particular, the Lorentzian part of the Hamiltonian is given by an operator representing the scalar curvature of the spatial manifold \cite{Lewandowski:2021iun, Lewandowski:2022xox}.

The time evolution of an initial state generated by the physical Hamiltonian is computed numerically. The task of evaluating the action of the time evolution operator on a given state vector is made numerically accessible by introducing a truncation of the Hilbert space of the model. This truncation is achieved by imposing a finite but relatively large cutoff on the spin quantum numbers. In this way the problem is converted into a calculation on a finite-dimensional Hilbert space, which can be performed by making use of functions provided by standard numerical libraries for efficient computation of the action of the exponential of a sparse matrix on a vector. Throughout the evolution of the state vector, we monitor the occupation number of the basis states in which any spin is equal to the cutoff value, using this as a rough criterion for detecting when the computation carried out on the truncated Hilbert space has reached the end of its range of validity.

We considered the time evolution of various initial states, peaked on classical data corresponding to a static, an expanding or a contracting universe, in two different models: one where the physical Hamiltonian operator consists of the Euclidean part only, and another where the physical Hamiltonian is given by the full Lorentzian expression. The quantum dynamics of the state, as described by the evolution of the expectation values of the volume operator and the holonomy operator, is compared against the semiclassical effective dynamics of a homogeneous and isotropic universe. In most cases we find that the dynamics of the quantum observables closely follows the semiclassical trajectory, at least over the time interval in which the evolution of the quantum state is not yet affected by the presence of the cutoff.

In the case of states describing an initially contracting universe, our computations confirm that the contraction is halted by the quantum dynamics and the state enters an expanding phase after going through a ``bounce''. This is of course not an unexpected result, as the phenomenon of a classical singularity being resolved by quantum effects and becoming replaced with a bounce in the quantum theory is a very well established phenomenon in the literature of loop quantum cosmology and quantum models of spherically symmetric spacetimes \cite{Li:2023dwy, Gambini:2022hxr}. In terms of comparing the quantum evolution with the effective semiclassical dynamics, we encountered some examples of initially contracting states in which the quantum dynamics begins to deviate from the semiclassical trajectory already at a relatively early stage of the evolution, where the numerical computation can still be expected to accurately describe the true dynamics of the state (according to the criterion based on the occupation number of basis states adjacent to the cutoff). This discrepancy can at least tentatively be attributed to the fact that the ``poorly behaved'' examples tend to be those where the evolution reaches a relatively low value of the volume before turning into an expansion. Hence it seems possible that the spins of the basis states which are relevant to the dynamics of such states near the bounce are simply too small to be compatible with the assumption of large spins that is inherent to the kinematical framework of quantum-reduced loop gravity.

During our attempts to gain some insight into the behavior described above, we analyzed a series of examples, in which the dynamics of the same initial state is computed under different (artificial, unphysical) variants of the Euclidean part of the physical Hamiltonian. These calculations show that there seems to be a direct connection between the deteriorated semiclassical behavior of some of the contracting initial states, and the presence of reduced flux operators raised to negative powers in the physical Hamiltonian. In all examples where the Hamiltonian is free of negative powers of the flux operators, the evolution of a particular initial state follows the semiclassical trajectory extremely closely, and the state seems to fully retain its semiclassical properties through the bounce (\Figs \ref{fig:H_abc-good-1} and \ref{fig:H_abc-good-2}). In contrast, the evolution of the same state under the actual Euclidean Hamiltonian of the one-vertex model begins to noticeably diverge from the semiclassical trajectory already in the region around the bounce (\Fig \ref{fig:H_E-bounce-2}).

In order to better understand this behavior and its possible physical significance, it would be important to further extend our computations to other physically relevant regularizations of the Hamiltonian, and not only to unphysical toy examples of the type considered in section \ref{sec:H_abc}. The findings presented in section \ref{sec:H_abc} suggest that the discrepancy between the quantum dynamics and the semiclassical effective dynamics observed in some of the examples may be related to the Tikhonov regularization used to quantize the inverse volume element, since it is this regularization which is the source of the negative powers of the flux operator in the Hamiltonian. A natural candidate for an operator which does not rely on the Tikhonov regularization and is free of any negative powers of the fluxes would be Thiemann's well-known regularization of the Hamiltonian. While likely to be somewhat more challenging from the practical point of view, we expect that it would certainly be feasible to repeat the computations performed for this work under the dynamics generated by Thiemann's Hamiltonian adapted to the one-vertex model. The results would undoubtedly help to shed some light on whether the disagreement between the quantum and semiclassical dynamics is a behavior shared among several different regularizations of the Hamiltonian, or a feature specific to the Tikhonov regularization.

A comparison of the dynamics generated by a Thiemann-type Hamiltonian against the Hamiltonian used in this article would also be interesting from the point of view of deciding between the many different regularizations available for constructing the Hamiltonian operator in canonical loop quantum gravity. If it could be established that the use of the Thiemann regularization (or possibly some other alternative regularization of the Hamiltonian) consistently leads to better semiclassical properties of the dynamics and fails to display the kind of discrepancies observed for the Hamiltonian considered here, this could be taken as a clear, definite indication to prefer the alternative Hamiltonian over the one obtained using the Tikhonov regularization. Any concrete physical argument favoring some regularization of the Hamiltonian over another would of course be very welcome, as it would serve to at least somewhat limit the very great number of different possibilities that are currently available for the quantization of the Hamiltonian in loop quantum gravity.

\subsection*{Acknowledgments}

The author thanks Hanno Sahlmann and Waleed Sherif for helpful discussions, Mehdi Assanioussi for a careful reading of the manuscript, and the members of the quantum gravity group at the Centre de Physique Théorique in Marseille, in particular Pietro Donà, Alejandro Perez and Simone Speziale, for their hospitality during a long-term visit, where the idea for this work was initially conceived. This work was funded by National Science Centre, Poland through grant no.~2022\slash 44\slash C\slash ST2\slash 00023. For the purpose of open access, the author has applied a CC BY 4.0 public copyright license to any author accepted manuscript (AAM) version arising from this submission.

\subsection*{Data availability statement}

The full set of data analyzed in this work is published at the Zenodo repository \url{https://zenodo.org/records/20328115}. The data can be reproduced by means of the code available at the associated GitHub repository (see code availability statement below).

\subsection*{Code availability statement}

The source code of the programs used to perform the computations presented in this article is made available at the GitHub repository \url{https://github.com/imakinen/QRLG-evolution}. The repository contains an example script for reproducing the data used to create the plots included in this article.

\setlength\bibitemsep{0.5\baselineskip}
\printbibliography

\end{document}